\documentclass[11pt]{article}
\usepackage{setspace}
\usepackage{graphicx}
\usepackage{amsmath}
\usepackage{amssymb}
\newcommand{\pr}[1]{\mbox{\rm Pr}\left(#1\right)}

\newtheorem{corol}{Corollary}
\newtheorem{lemma}{Lemma}
\newtheorem{rmk}{Remark}
\newtheorem{thm}{Theorem}
\newcommand{\N}{\mathbb{N}}

\newcommand{\dff}{\mbox{\it dif}}

\title{Key Distribution Protocols Based on Extractors Under the Condition of Noisy Channels in the Presence of an Active Adversary}
\author{Viktor~Yakovlev, 
  Valery~Korzhik, 
  Guillermo Morales-Luna
  and Mihail Bakaev
\thanks{V. Yakovlev, V. Korzhik and M. Bakaev are with the Department of Information Security of Telecommunication Systems, State University of Telecommunication, St. Petersburg, Russia e-mail: viyak@bk.ru, korzhik1@bk.ru.}
\thanks{G. Morales-Luna is with the Computer Science Department, CINVESTAV-IPN, Mexico City, Mexico e-mail: gmorales@cs.cinvestav.mx. Dr. Morales-Luna acknowledges the partial support of Mexican CONACyT.}}

\begin{document}
\maketitle

\begin{abstract}
We consider in this paper the information-theoretic secure key distribution problem over main and wire-tap noise channels with a public discussion in presence of an active adversary. In contrast to the solution proposed by ourselves for a similar problem using hashing for privacy amplification, in the current paper we use a technique of extractors.

We propose modified key distribution protocols for which we prove explicit estimates of key rates without the use of estimates with uncertain coefficients in notations $O,\Omega,\Theta$.

This leads in the new conclusion that the use of extractors is superior to the use of hash functions only with the very large key lengths $\ell$ (of order $\ell>10^5$ bits).

We suggest hybrid key distribution protocols consisting from two consecutively executed stages. At the fist stage it is generated a short authentication key based on hash function, whereas at the second stage it is generated the final key with the use of extractors. We show that in fact the use of extraction procedure is effective only at the second stage. We get also some constructive estimates of the key rates for such protocols.

\end{abstract}


\section{Introduction}

Advances in design and implementation of quantum computers~\cite{i} as well as design of super-fast multiprocessor conventional computers threat some conceptually secure cryptosystems. Hence perfect one-time pad ciphers proposed by Shannon~\cite{ii} are necessary. But the use of perfect ciphers requires key lengths proportional to messages~\cite{iii}. This inconvenience can be solved with the use of key distribution over communication channels protected from eavesdropping. There are several approaches in order to remove (or at least to control) an eavesdropping on the keys:
\begin{itemize}
 \item quantum channels~\cite{iv},
 \item methods based on fluctuation of radio wave channels~\cite{v,vi,vii,viii},
 \item Wyner's concept of wire-tap channel,
 \item key generation by hashing of random string initially distributed over noisy channels~\cite{xiii,xiv,xv,xvi,xvii,xviii,xix,xx,xxi,xxii,xxiii,xxiv,xxv,xxvi}.
\end{itemize}

In the current paper, we follow the last approach. The most advanced results in this setting, under the condition of an active adversary have been obtained by Maurer and Wolf. They proposed several key distribution protocols~\cite{xv,xvi,xvii,xviii,xix,xx,xxi,xxii,xxiii} and made a performance comparison of asymptotic and non-asymptotic key rates for a given level of key security.

We considered in~\cite{xxvi} some modification of the Maurer and Wolf's MW-protocol consisting in using an authentication algorithm over noisy channels, called by ourselves the {\em $\alpha$-protocol}, instead of the request-response algorithm presented in~\cite{xxi}. In the same paper~\cite{xxvi}, we proposed also the {\em $\beta$-protocol} that differs from the $\alpha$-protocol in absence of the hash function transmission over public discrete channel because the hash function can be formed from the string which the users have got just after the execution of the initialization phase. Using the $\beta$-protocol entails an increasing of the key rate in several cases. We proposed also in~\cite{xxvi} the so called $\alpha'$ and {\em $\beta'$-protocols} in which special initially distributed short keys are used in order to provide authentication procedures over {\em public discussion channels} (PDC).

Hybrid protocols comprising pairs of sequentially executed protocols $(\alpha,\alpha')$, $(\alpha,\beta')$, $(\beta,\alpha')$, $(\beta,\beta')$ were investigated in~\cite{xxvi}. The first protocol in each pair is used to generate an authentication key, whereas the second one provides a generation of the main secret key for encryption/decryption given the authentication key. The relation among the key rates and a comparison of protocol performance evaluation were also introduced.

The main feature of the protocols considered in~\cite{xxvi} is their strict constructiveness because the parameters determining their efficiency do not contain unknown coefficients typical for $O,\Omega,\Theta$-estimations.

Our contribution and novel content in the current paper are the following:
\begin{enumerate}
\item We propose some new (modified) key distribution protocols using extractors. We prove explicit estimates of key rates without the use of estimates of uncertain coefficients in $O,\Omega,\Theta$-estimations. (In~\cite{xxvi} we solved the similar problem using hash functions instead of extractors).

In contrast to~\cite{xxi}, we consider a scenario where the legal users are able to receive raw bit strings over noisy channels and as a consequence they are pairwise distinct. This entails the need to send check symbols from user A to user B in order to agree the raw bit strings received by legal users. By the same reason, we have changed the authentication algorithm: instead of a request-response algorithm~\cite{xxi}, we use a non-interactive one based on the authentication code.
 
A consideration of the non-asymptotic case leads us in the new conclusion that the use of extractors is superior to the use of hash functions only for very large key lengths ($\ell$) of the order of $10^5$ bits.
\item We suggest hybrid key distribution protocols consisting of two consecutively executed stages. At the first stage, a short authentication key based on a hash function is generated, whereas at the second stage, the final key using extractors is generated. We show that in fact the use of an extraction procedure is effective only at the second stage. We get also explicit estimates of key rates for such protocols.
\item We prove also an asymptotic behavior of the key rates for all considered protocols that allows to compare the potential efficiency of them with the potential efficiency of protocols considered here and in~\cite{xxvi}.
\end{enumerate}
The outline of this paper is the following: Section~\ref{sc.02} contains the preliminaries and descriptions of the main procedures to be used in key distribution protocols. In Section~\ref{ssc.02} we describe the model of key distribution based on noisy wire-tap channels in the presence of an active adversary and we introduce the main criteria for key distribution protocol efficiency. We introduce main procedures as error correction, authentication and privacy amplification (based both on hashing and extraction). In section~\ref{sc.03} we describe the $\alpha_{ext}$-protocol, and the new key distribution $\beta_{ext}$-protocol without transmission of the extractor's seed on the public discussion channel and we prove their main features. In section~\ref{sc.04} we present a modification of the previous $\alpha'_{ext}$ and $\beta'_{ext}$-protocols under the condition that initially the legal users share short authentication keys. In section~\ref{sc.05} we describe the so called hybrid protocols as combinations of different pairs of single protocols and we estimate their performance evaluation. In section~\ref{sc.06} we conclude the paper.

\section{Main notions and procedures involved in the key distribution protocol} \label{sc.02}

Here, we repeat mostly the content of the same point as in~\cite{xxvi}. It is done in order to provide an independent reading of the current paper.

\subsection{Model for key distribution and the main criteria for protocol efficiency} \label{ssc.02}

Let us consider the model of key distribution between a legal user, {\em Alice} (A), and another user, {\em Bob} (B), in the presence of an active adversary, {\em Eve} (E), assuming that initially the legal users do not have shared secret keys. The {\em key distribution protocol} (KDP) consists of two phases: {\em initialization} and {\em key generation}.

\begin{figure}[!t]
\centering
\includegraphics[width=2.5in]{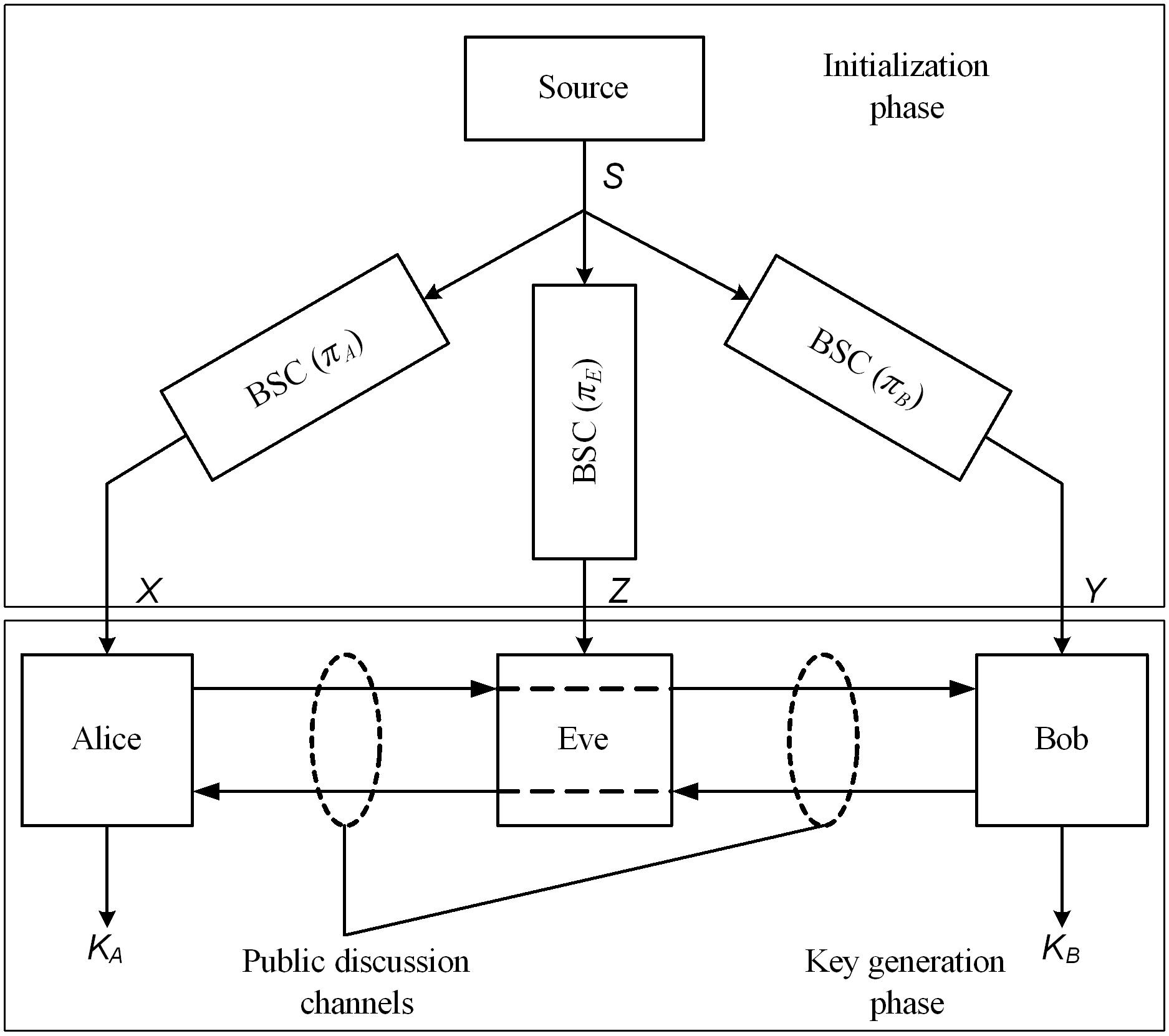} \hspace{1em} 
\caption{Model of key distribution protocol over noisy legal channels in presence of an active adversary.}
\label{fig.01}
\end{figure}

In the KDP initialization phase, A, B, and E receive random i.i.d. sequences $X=\left\{x_i\right\}_{i=1}^k$, $Y=\left\{y_i\right\}_{i=1}^k$, $Z=\left\{z_i\right\}_{i=1}^k\in\{0,1\}^k$, respectively, such that for each $i$, $\pr{x_i\not=y_i} = p_m$ and $\min\{\pr{x_i\not=z_i},\pr{y_i\not=z_i}\} = p_w$ (see Figure~\ref{fig.01}). 
One of the methods to provide legal users A, B with the sequences $X$, $Y$ is to generate the truly random sequence $S=\left\{s_i\right\}_{i=1}^k\in\{0,1\}^k$ by some trusted party, say {\em source} S, and then to transmit it to the legal users A and B over noisy channels (as in the {\em source model}~\cite{xiii},~\cite{xvi}). We will assume that A and B receive the sequences $X$, $Y$ over {\em binary symmetric channels} (BSC) without memory with error probabilities $\pi_A=\pr{x_i\not=s_i}$, $\pi_B=\pr{y_i\not=s_i}$, while the adversary E receives the sequence $Z$ over a BSC with error probability $\pi_E=\pr{z_i\not=s_i}$. It is easy to see that if the original sequence $S$ is truly random then the same property holds for the sequences $X$, $Y$ and $Z$. (Examples of practical implementation of the initialization phase in real world can be found in~\cite{xxvi}). In this phase it is natural to assume that the adversary is unable to intervene the transmission from S to A and B. 

The key generation phase consists in an information exchange over a {\em public discussion channel} (PDC) with a goal to share eventually the final key. We note that the use of PDC is necessary in order to send check symbols to test the agreement of the strings X and Y and sometimes for the parameters of the hash function or extractor seed transmission (see details in the following sections). The adversary E can receive all information transmitted over the PDC. We assume also that the PDC's between legal users and E are binary noiseless channels (if E does not intervene in transmission). However E can change or replace this information as desired and therefore it is necessary to authenticate messages transmitted over a PDC in order to detect any intervention of E and to reject suspicious messages.

Let us define the following parameters of the key distribution protocol characterization:
\begin{description}
\item[{\rm $\ell$:}] the key length (the number of bits which are contained in the keys $K_A$ and $K_B$),
\item[{\rm $I(K_A, U)$:}] \hspace{.75cm}the amount of Shannon's information in possession of the adversary E about the final key $K_A$ after receiving all acceptable information $U$, including the sequence $Z$ and the other messages transmitted over the PDC
\item[{\rm $P_e=\pr{K_A\not=K_B}$:}] \hspace{2.25cm}the probability of legal users keys disagreement,
\item[{\rm $P_f$:}] the probability of false rejection of the KDP protocol (when A or B falsely believe that E has intervened the PDC),
\item[{\rm $P_d$:}] the probability of deception false information provided by E during information transmission over PDC (it can result in an opportunity to fix a key between any legal user and E although leaving the legal user on the belief that he (she) has shared a key with his (her) legal partner),
\item[{\rm $R_k$:}] the key distribution rate (the ratio of the key length $\ell$ to the length of sequences $X$, $Y$),
$
R_k = \frac{\ell}{k}. 
$
\end{description}
It is reasonable to impose the following conditions on the KDP:
\begin{eqnarray}
\ell &=& \ell^{req}, \label{eq.002} \\
I(K_A, U) &\leq& I^{adm}, \label{eq.003} \\
P_e &\leq& P_e^{adm}, \label{eq.004} \\
P_f &\leq& P_f^{adm}, \label{eq.005} \\
P_d &\leq& P_d^{adm}, \label{eq.006} 
\end{eqnarray}
where $\ell^{req}$ denotes the {\em required key length} and the superscript {\em adm} stands for {\em admissible parameter value}. We will say that the above conditions are {\em requirements of the KDP}. The efficiency of the KDP will be estimated by the key rate $R_k$ and then among all protocols satisfying~(\ref{eq.002})-(\ref{eq.006}), we will select the most efficient by making $R_k$ to attain its largest value. As we will show later, some inequalities~(\ref{eq.003})-(\ref{eq.006}) may randomly hold. Then an additional requirement can be stated as
\begin{equation}
P_{risk} \leq P_{risk}^{adm}, \label{eq.007}
\end{equation}
where $P_{risk}^{adm}$ is the probability that at least one of the inequalities~(\ref{eq.003})-(\ref{eq.006}) does not hold. 

\subsection{Known asymptotic results regarding key rates}

Let us denote by $R^*$, $R^{**}$ the maximum achievable key rates in a KDP between the legal users under the condition of a passive or active adversary, respectively. In the papers~\cite{xvi},~\cite{xviii},~\cite{xxi},~\cite{xxiii} the proofs of these values were presented. For the source model of the wire-tap channel with initialization phase in the KDP using BSC with probabilities $\pi_A$, $\pi_B$, $\pi_E$ the following theorem holds:

\begin{thm}[see~\cite{xxiii}]\label{th.01} If $\pi_E > \pi_A$ and $\pi_E > \pi_B$ , then $R^* = R^{**}$. If either $\pi_E \leq \pi_A$ or $\pi_E \leq \pi_B$ then $R^{**} = 0$.
\end{thm}

We note that under the conditions $\pi_E > \pi_A$ and $\pi_E > \pi_B$, the users A and B either share the key or they may detect interception in the case of E's intervention. This fact cannot be interpreted as a defect of KDP because E can use even a simple strategy: she tries to break off the PDC between legal users in order to impede the completion of the KDP. Let $p_m$, $p_w^A$, $p_w^B$, denote the probabilities of disagreements among the sequence pairs $(X,Y)$, $(X,Z)$, $(Y,Z)$ respectively. Then
$$
\begin{array}{rcl}
p_m &=& \pi_A + \pi_B - 2 \pi_A \pi_B \\
p_w^A &=& \pi_A + \pi_E - 2 \pi_A \pi_E \\
p_w^B &=& \pi_B + \pi_E - 2 \pi_B \pi_E 
\end{array} 
$$
It is easy to see that if $\pi_E > \pi_A$ then $p_w^A>p_m$ and similarly if $\pi_E > \pi_B$ then $p_w^B>p_m$. We will consider the worst case for legal users as $p_w=\min\{p_w^A,p_w^B\}$.

After the execution of the initialization phase the source model is reduced to the {\em channel model} where user A sends the sequence $X^k$ to user B who receives it as $Y^k$, whereas E receives $X^k$ as $Z^k$. Then the probability of error on the {\em main virtual} BSC between A and B is $p_m$ and the probability of the {\em wire-tap virtual channel} from A to E is $p_w$. (The PDC remains the same after such reduction of the source model to the channel model.)

\begin{thm}[see~\cite{xxiii}]\label{th.02} In the channel model setup with probabilities $p_m$, $p_w$ the maximum key distribution rate is 
\begin{equation}
R^* = g(p_w) - g(p_m), \label{eq.009}
\end{equation}
where
$
p\mapsto g(p) = -p\,\log p -(1-p)\,\log (1-p) 
$ 
is the {\em entropy function}.
\end{thm}

\subsection{Error correcting codes}

Let $C$ be a binary linear error-correcting $(k+r,r)$-code and let $C^r$ be a string consisting of $r$ check symbols. It has been proved in~\cite{xxx} that if the information symbols are transmitted on the BSC with the error probability $p_m$, whereas the check symbols on the noiseless channel, then the average error probability of decoding on the ensemble of all $(k+r,r)$-codes meets the following modified Gallager's bound
\begin{equation}
P_e\leq 2^{-k\,E(R_c)}, \label{eq.011}
\end{equation}
where
\begin{eqnarray}
E(R_c) &=& \max_{\rho\in(0,1)}\left[E_0(\rho) - \frac{\rho(2R_c-1)}{R_c}\right], \label{eq.012} \\
E_0(\rho) &=& \rho - (1+\rho)\,\log_2\left(p_m^{\frac{1}{1+\rho}} (1-p_m)^{\frac{1}{1+\rho}}\right) \label{eq.013} 
\end{eqnarray}
is Gallager's function for a BSC with the error probability $p_m$
\begin{equation}
R_c = \frac{k}{k+r} \label{eq.014}
\end{equation}
is the code rate. We note that in the frame of the above model, the code rate $R_c$ satisfies the inequality
\begin{equation}
0\leq \frac{2R_c-1}{R_c} \leq C^*, \label{eq.015}
\end{equation}
where $C^* = 1 - g(p_m)$ is the capacity of the BSC with the probability of error $p_m$.

It follows from~(\ref{eq.015}) that $\frac{1}{2} \leq R_c \leq \frac{1}{1+g(p_m)}$. In the asymptotic case $R_c\to \frac{1}{1+g(p_m)}$, then
\begin{equation}
r = k\,g(p_m). \label{eq.016}
\end{equation}
We see from~(\ref{eq.016}) that an arbitrary small value of the erroneous decoding probability is achieved for large block length if the number $r$ of check symbols (but not block length) is proportional to the number of information symbols $k$ with coefficient $g(p_m)$.

\subsection{Authentication based on the class of universal hash functions}

In order to execute the authentication procedure, we use the universal hash function which are described below.

For any finite set $A$, let $|A|$ denote its cardinality. For any two finite sets $A, B$, let $H$ be a set of hash-maps $A\to B$. For each $x_0, x_1\in A$, let $\delta_H(x_0, x_1) = \left|\{h\in H|\ h(x_0) = h(x_1)\}\right|$ be the number of hash functions in $H$ that collide in $x_0$ and $x_1$. We recall that $H$ is {\em universal}$_2$, $U_2$ in short, if for each $x_0, x_1\in A$, $\delta_H(x_0, x_1) \leq \frac{|H|}{|B|}$.

Let $P_{col}$ be the so called {\em collision probability}, namely the probability that there occurs a pair of elements in $A$ colliding under an uniformly chosen map $h\in H$. Clearly, $P_{col}\leq |B|^{-1}$. The class $H$ is {\em strongly universal}$_2$, $SU_2$, if
\begin{equation}
\forall x\in A,\ \forall b\in B:\ \ |\{h\in H|\ h(x)=y\}| = \frac{|H|}{|B|} \label{eq.016a}
\end{equation}
and besides for any distinct $x_0, x_1\in A$, and any $y_0, y_1\in B$,
$$|\{h\in H|\ h(x_0)=y_0\ \&\ h(x_1)=y_1\}| \leq \frac{|H|}{|B|}.$$
For a given $\epsilon>0$, the class $H$ is {\em $\epsilon$-almost universal}, $\epsilon$-$AU_2$, if for all $x_0, x_1\in A$: $\delta_H(x_0, x_1) \leq \epsilon\,|H|$. The class $H$ is {\em $\epsilon$-almost strongly universal}, $\epsilon$-$ASU_2$, if~(\ref{eq.016a}) holds and for any pairs $x_0, x_1$ and $y_0, y_1$ of distinct points in $A$ and $B$,
$$|\{h\in H|\ h(x_0)=y_0\ \&\ h(x_1)=y_1\}| \leq \epsilon\frac{|H|}{|B|^2}.$$
Naturally, each class $|B|^{-1}$-$ASU_2$ is also $SU_2$.

{\em Examples of hash functions classes}: We assume that the sets $A$ and $B$ consist of all binary sequences of lengths $a$ and $b$, respectively: $A = \{0,1\}^a$, $B = \{0,1\}^b$, hence $|A| = 2^a$, $|B| = 2^b$.
\begin{description}
\item[An $U_2$ class.] \hspace{1cm}The set $A$ can be identified with the Galois field $GF(2^a)$. For each $s\in GF(2^a)$, let $h_s: A \to B$, $x\mapsto\lfloor xs\rfloor_b$, where the map $z\mapsto\lfloor z\rfloor_b$ takes the $b$ least significant bits in $z$. The class $\left\{h_s\right\}_{s\in A}$ is $U_2$. Such hash functions are described uniquely by binary strings of length $a$.
\item[An $SU_2$ class.] \hspace{1.1cm} For each $s,t\in GF(2^a)$, let $h_{st}: A \to A$, $x\mapsto sx + t$. The class $\left\{h_{st}\right\}_{s,t\in A}$ is $SU_2$ and clearly this class can be indexed by sequences of length $2a$.
\item[An $\epsilon$-$ASU_2$ class.] \hspace{1.6cm} It has been shown in~\cite{xxxi} that the hash functions chosen from an $\epsilon$-$ASU_2$ class are connected with {\em incomplete balanced schemes}. The parameters of the $\epsilon$-$ASU_2$ class can be described as 
\begin{equation}
|A| = q^{2^i}\ ,\ |B| = q\ ,\ |H| = q^{i+2}\ ,\ \epsilon = \frac{i+1}{q}, \label{eq.017}
\end{equation}
where $q$ is a power of a prime and $i>1$ is an integer.
\end{description}
Let us analyze the procedure of message authentication. Let $x$ be the message to be authenticated during its transmission from user A to user B. User A forms the authenticator $y=h(x)$ of his message $x$ using the keyed hash function $h\in H$ known by him (but unknown for adversary the E), then A appends $y$ to $x$ and sends the pair $(x,y)$ to the legal user B. In order to check the authenticity of the message $x$, the user B receives a pair $(\tilde{x},\tilde{y})$ (which may be forged), B forms the authenticator $\tilde{\tilde{y}}=h(\tilde{x})$ with his knowledge of the secret hash function $h$ and compares $\tilde{\tilde{y}}$ with $\tilde{y}$. If they coincide then B accepts $x$, otherwise he rejects it.

It was shown in~\cite{xxxi} that if the hash functions, chosen from the $\epsilon$-$ASU_2$ class, are used in the authentication procedure then for the best adversary's strategy consisting in an impersonation or substitution of the messages, the following probability bounds hold
\begin{eqnarray}
P_i &\leq& |B|^{-1}, \label{eq.018} \\
P_s &\leq& \epsilon, \label{eq.019} 
\end{eqnarray}
where $P_i$ is the probability of message impersonation, and $P_s$ is the probability of message substitution.

Let us define the probability of {\em undetected false message deception} by the adversary as $P = \max\{P_i,P_s\}$. The bounds~(\ref{eq.018}),~(\ref{eq.019}) will hold only if the active adversary ignores completely the used hash function $h$ in the authentication procedure. But there may be situations when the keyed hash function is partly known by the adversary although authentication procedure is still possible. In order to clarify this situation let us recall initially from~\cite{xviii} that for a discrete random variable $\xi$ taking values over a set $X$ with probability distribution $P_{\xi}$ the minimal entropy is
$$
H_{\infty}(\xi) = -\log\max_{x\in X}P_{\xi}(x), 
$$
and the Renyi entropy of the random variable $\xi$ is
\begin{equation}
H_2(\xi) = -\log\sum_{x\in X}P_{\xi}^2(x). \label{eq.021}
\end{equation}

\begin{thm}[see~\cite{xxxviii}]\label{th.03} Suppose legal users A and B have the random key $h$ with length $\ell_0$ within an authentication scheme based on $\epsilon$-$ASU_2$ hash functions where $\epsilon = 2^{-\tilde{b}}$. Denote by $U$ the total knowledge of E about $h$. Then, assuming that for any sample $u$ 
\begin{equation}
H_{\infty}(h\,|\,U=u) \geq t\ell_0,\ \ \ \ 0<t<1, \label{eq.021a}
\end{equation}
the probability $P_d$ of {\em message undetected deception} is upper bounded as
$$
P_d \leq 2^{-\left(\frac{\tilde{b}-\ell_0(1-t)}{2}-1\right)}. 
$$
\end{thm}

\subsection{Authentication based on noisy channels}\label{ssc.auth}

The message authentication considered above and based on the use of hash functions from either the class $SU_2$ or the class $\epsilon$-$ASU_2$ requires a possession by legal users of the secret or partly secret keys. However such keys cannot be taken directly from the strings $X^k$, $Y^k$ shared in the initialization phase because they differ even for legal users. On the other hand it is impossible to conciliate these string by sending from A to B the check symbols strings of $X^k$ because PDC is get not authenticated and B could ``conciliate'' formerly the false string $Z^k$ with E. 

In order to avoid this situation it is necessary firstly to design a {\em keyless message authentication} based on noisy channels. In~\cite{xvi} a special type of codes has been proposed in order to solve this problem: the so called {\em authentication codes} (AC). Let us describe them briefly.

In an initialization phase the users share the strings $X^k$, $Y^k$ over a BSC ($\pr{x_i\not=y_i} = p_m$) and they agree an error correcting binary systematic $(n_a,k_a)$-code $V$ in order to authenticate a length $k_a$ message. The authenticator ${\bf w} = (w_1,\ldots,w_{n_a})$ of a message {\bf m} is formed as follows: for each $i$ let $v_i$ be the $i$-th bit of the codeword in $V$ corresponding to {\bf m} and $w_i=v_i$ if $v_i=1$, or let it remain undefined otherwise.

After receiving a pair $(\tilde{{\bf m}},\tilde{{\bf w}})$, the user B forms his authenticator $\tilde{\tilde{{\bf w}}}$ for the message $\tilde{{\bf m}}$ using his string $Y^k$ according to the agreed procedure and compares $\tilde{{\bf w}}$ with $\tilde{\tilde{{\bf w}}}$. If the number of the coinciding bits in them is less or equal to some given threshold $\Delta_w$ then the message $\tilde{{\bf m}}$ succeeds as authentic, otherwise it is removed as forged. The AC's were investigated in~\cite{xvi} and can be characterized by two probabilities:
\begin{description}
\item[$P_f$.] the probability of false removal of the message although adversary E does not intervene at all;
\item[$P_d$.] the probability of the deception of false message, i.e. the probability of the event that E has forged a message and this fact was not detected by B.
\end{description}
$P_f$ and $P_d$ do not depend on ordinary minimum code distance of the code $V$ but on the so called {\em minimum asymmetric semidistance} $d_{01}$ that is determined by the minimal number of differences between 0 and 1 symbols in any pair of distinct code words of $V$.

\begin{thm}[see~\cite{xxvi,xxxii}]\label{th.04} Let $V$ be an $(n_a,k_a)$-AC with constant Hamming weight $\tau$ for all non-zero codewords and with asymmetric semidistance $d_{01}$. Then the probabilities $P_f$ and $P_d$ for the authentication procedure on noisy wire-tap channel with parameters $p_m$ and $p_w$, can be upper bounded as follows:
\begin{eqnarray*}
P_f &\leq& \sum_{i=\Delta_w+1}^{\tau} {\tau\choose i}\,p_m^i(1-p_m)^{\tau - i}, \\ 
P_d &\leq& \sum_{i=0}^{\Delta_w} {d_{01}\choose i}\,p_m^i(1-p_m)^{d_{01} - i}\cdot \\ 
 &&\hspace{2em} \sum_{j=0}^{\Delta_w-1} {{\tau-d_{01}}\choose j}\,p_m^j(1-p_m)^{\tau-d_{01} - j}. 
\end{eqnarray*}
\end{thm}
It is a very hard problem to find $d_{01}$ for any linear code. But there exists a very simple method to design the code $V$ with known $d_{01}$, given the linear $(n_0,k_0)$-code $\tilde{V}$ with known ordinary minimum code distance $d$ proposed in~\cite{xvi}.

Namely, let us substitute the symbol {\tt 1} with the symbol pair {\tt 10} and the symbol {\tt 0} with {\tt 01} in $\tilde{V}$.Then evidently the parameters of the code $V$ are: 
\begin{equation}
n_a = 2 n_0\ \ ,\ \ k_a = k_0\ \ ,\ \ d_{01}=d\ \ ,\ \ \tau = n_0. \label{eq.025}
\end{equation}
We have proved in~\cite{xxvi} the following theorem with the use of the above code.

\begin{thm}\label{th.05} Let $V$ be a $(k_0+r_0,k_0)$-error correction code with minimum distance $d$ that is used in the authentication procedure. Then for any $p,q>0$ there exists an integer $k_0'$ and an AC, guaranteeing $\frac{r_0}{k_0} <q$, $P_f\leq p$, $P_d\leq p$ for all $k_0>k_0'$.
\end{thm}
It follows from this theorem that
$$
\frac{r_0}{k_0+r_0} \to 0 \mbox{ as }k_0\to +\infty.
$$
This means that the length of the authenticator approaches zero as the block length tends to infinity. Other methods to design constant weight AC were investigated in~\cite{xxxiii}. 

\subsection{Extractors}\label{ssc.ext}

Let us recall the notion of {\em extractor} and {\em strong extractor}~\cite{xxxv,xxxvi,xxxvii}.
 Two probability distributions $P, Q$, defined on the same set $X$, are called {\em $\epsilon$-close} if their statistical difference
$$
\dff (P,Q) = \frac{1}{2} \sum_{x\in X}|P_X(x)-Q_X(x)| 
$$
does not exceed $\epsilon$.
A map $E:\{0,1\}^k\times\{0,1\}^u\to\{0,1\}^{\ell}$ is an {\em $(\eta,\epsilon)$-extractor} if for any probability distribution random variable $X$ on $\{0,1\}^k$ such that $H_{\infty}(X)\geq\eta$ and any uniformly distributed random variable $\Gamma$ on $\{0,1\}^u$, the statistical difference probability distribution of the extractor output $E(X,\Gamma)$ with respect to an uniform distribution on $\{0,1\}^{\ell}$ is at most $\epsilon$. In order words, the extractor maps a random sequence $X$ of length $k$ with symbols taken from an ensemble of minimal entropy $H_{\infty}(X)$ to a random sequence of length $\ell$ that is $\epsilon$-close to an uniformly distributed sequence with the help of a truly random sequence $\Gamma$ of length $u$. The last sequence can be seen as a ``seed'' of the extractor.
The extractor $E(X,\Gamma)$ has parameters $(k,\eta , u, \ell, \epsilon)$, where $k$ is the length of input random sequence, $\eta$ is the evaluation of minimal entropy ($H_{\infty}(X)\geq\eta$) on the set of input sequences, $u$ is the length of the seed $\Gamma$, $\ell$ is the length of the output sequence, and $\epsilon$ is the statistical distance between the output probability distribution and an uniform distribution on the output set.

A mapping $E:\{0,1\}^k\times\{0,1\}^u\to\{0,1\}^{\ell}$ is called a {\em strong extractor} $E(X,\Gamma)$ if for any probability distribution random variable $X$ on the set $\{0,1\}^k$ having minimal entropy $H_{\infty}(X)\geq\eta$ and for any uniformly distributed random variable $\Gamma$ on the set $\{0,1\}^u$ the probability distribution of the concatenated variables ($\Gamma\circ E(X,\Gamma)$) is close to an uniform distribution on $\{0,1\}^{\ell+u}$. More specifically 
$$
\dff \left(\Gamma\circ E(X,\Gamma),U^{u+\ell}\right) \leq \epsilon. 
$$
This means that the strong extractor provides the closeness of probability distribution for the concatenation of the output extractor sequence and the seed sequence to an uniform distribution. In the current paper, we will consider only extractors based on the construction~\cite{xxxv,xxxvi} which is an improvement of the originally proposed by Trevisan~\cite{xxxvii}. 

\begin{thm}[see theorem 22 in~\cite{xxxvi}]\label{th.06} For every $k$, $H_{\infty}(X)$, $\ell\in\N$ and $\epsilon > 0$, such that $\ell \leq H_{\infty}(X) \leq k$, there are explicit strong $(H_{\infty}(X), \epsilon)$-extractors $E:\{0,1\}^k\times\{0,1\}^u\to\{0,1\}^{\ell-\Delta}$ with
\begin{equation}
u=O\left(\frac
{\log_2^2\left(\frac{k}{\epsilon}\right)}
{\log_2 \left(\frac{H_{\infty}(X^k)}{\ell}\right)}\right), \label{eq.028}
\end{equation}
or
\begin{equation}
u= O\left(\log_2^2\left(\frac{k}{\epsilon}\right)\right)\cdot\log_2\left(\frac{1}{\mu}\right), \label{eq.029}
\end{equation}
where $1+ \mu= \frac{k}{\ell-1}$, $\mu < \frac{1}{2}$ and $\Delta=O(d)$. The value $\Delta$ is the {\em loss of extractor output sequence length}.
\end{thm}

The first extractor~(\ref{eq.028}), with $\frac{H_{\infty}(X^k)}{\ell}$ constant, is used for extraction of an arbitrary part of randomness ($H_{\infty}(X^k)$) from the input sequence $X^k$, whereas the second one~(\ref{eq.029}) is needed in order to extract all randomness $\ell = H_{\infty}(X^k)$ from the input sequence $X^k$. 

We are not going to use the estimates based on the $O$-operator and therefore let us find a more accurate estimate for the length of the seed. For this reason we consider in greater detail the design of the Trevisan's extractor modified by Raz, Reingold, Vadhan~\cite{xxxvi}.

\begin{figure}[!t]
\centering
\includegraphics[width=3.3in]{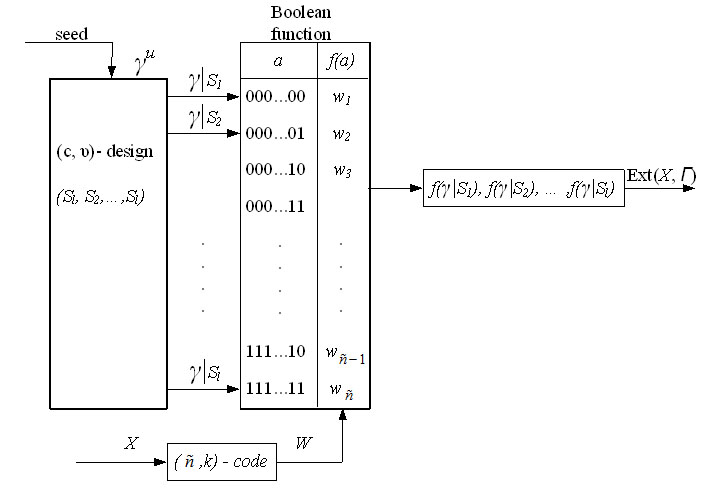} \hspace{1em} 
\caption{Design of the Trevisan's extractor.}
\label{fig.02}
\end{figure}

In order to design the Trevisan's extractor it is necessary to realize three components:
\begin{enumerate}
\item The linear error code $W$: With parameters $(\tilde{n},k)$ and minimal code distance $d_w$, where $\tilde{n}=2^{\nu}$, $\nu\in\N$. It is proposed to take this code as a concatenation of the Reed-Solomon and the Adamar codes.
\item Combinatorial block design scheme. ({\em Balance incomplete block design}, BIBD). This is a family of sets $S = \{S_1, S_2, \ldots,S_{\ell}\}$ holding the following properties:
\begin{eqnarray}
 & & S_i\subseteq\{1,2,\ldots,u\}, \nonumber \\ 
 & & |S_i| = \nu, \nonumber \\ 
i\not=j &\Longrightarrow& |S_i\cap S_j| < \log c\ \ \mbox{ with } c\geq 1. \label{eq.032} 
\end{eqnarray}
This means that the family consists of $\ell$ sets or blocks, each consisting of $\nu$ elements taken from the set of integers $\{1,2,\ldots,u\}$, while the number of elements contained simultaneously in any pair of blocks is at most $\log c$. Such construction is designated as a {\em $(\nu, c)$-scheme}.
\item Boolean function $f$: This map is defined over $\{0,1\}^{\nu}$ and for each $a_1,\ldots,a_{\nu}\in\{0,1\}$, $f(a_1,\ldots,a_{\nu})$ is a codeword of the $(\tilde{n},k)$-code $W$.
\end{enumerate}
The design of the extractor based on the three components given above is presented in Figure~\ref{fig.02}. 

The algorithm is executed in the following stages:
\begin{enumerate}
\item The input sequence $\overline{x}$ is encoded as $\overline{w}$ with the error correcting $(\tilde{n},k)$-code $W$. The word $\overline{w}$ gives the value of the Boolean function $f:\{0,1\}^{\nu}\to\{0,1\}^{\ell}$.
\item The random sequence $\gamma$ of length $u$ determines the samples $\gamma|S_i$, consisting of $\nu$ symbols of $\gamma$ with the use of blocks $S_i$ belonging to the $(\nu,c)$-BIBD. This means $\gamma|S_i=\left[\gamma\right]_{s\in S_i}$.
\item Output ${\bf w}=f\left(\left(\gamma|S_i\right)_{i=1}^{\ell}\right)$ as the result of the extractor.
\end{enumerate}
In the modified extractor version at~\cite{xxxvi}, it was proposed to use a {\em $(\nu,c)$-weak scheme}, in which the condition~(\ref{eq.032}) is changed by the condition 
$$
\sum_{j<i} 2^{|S_j\cap S_i|} \leq c(\ell -1), 
$$
where $c$ is some constant, $c>1$. The length $\tilde{n}$ of the code $W$ is chosen in~\cite{xxxvi}, p. 106, according to the condition
$$
\log(\tilde{n}) = O(\log\frac{k}{\epsilon}). 
$$
Since $w$ is the output of a Boolean function with $\nu$ arguments, $\tilde{n}$ should be equal to $2^{\nu}$. Obviously this condition will be fulfilled if 
\begin{equation}
\nu = \left\lceil\log\frac{k}{\epsilon}\right\rceil, \label{eq.035}
\end{equation}
where $\lceil x\rceil$ is the ``ceiling'' of $x$ (the least integer greater or equal than $x$).

The characterization of strong extractor is determined by the following statements.

\begin{thm}[Proposition 10 in~\cite{xxxvi}]\label{th.07} If $S =(S_1,\ldots,S_{\ell})$ (with $S_i\subset\gamma$) is a weak $(\nu,c)$-design for
\begin{equation}
c = \frac{1}{\ell}\left[H_{\infty}(X^k) - 3\log\frac{\ell}{\epsilon} - u - 3\right], \label{eq.036}
\end{equation}
then $E:\{0,1\}^{k}\times\{0,1\}^{u} \to\{0,1\}^{\ell}$ is a strong $(H_{\infty}(X^k),\varepsilon)$-extractor.
\end{thm}

\begin{thm}[Lemma 15 in~\cite{xxxvi}]\label{th.08} For every $\nu,\ell\in\N$ and $c>\ell$, there exists a weak $(\nu,c)$-design $S =(S_1,\ldots,S_{\ell})$ (with $S_i\subset\gamma$) and
\begin{equation}
u = \left\lceil\frac{\nu}{\ln c}\right\rceil\cdot\nu. \label{eq.037}
\end{equation}
Moreover, such a family can be found in polynomial time {\it poly}$(\ell,u)$.
\end{thm}

\begin{thm}[Lemma 17 in~\cite{xxxvi}]\label{th.09} For every $\nu,\ell\in\N$ and $0<\mu<\frac{1}{2}$, there exists a weak $(\nu,1+\mu)$-design $S =(S_1,\ldots,S_{\ell})$ (with $S_i\subset\gamma$) with
$
u = O\left(\nu^2\cdot\log\frac{1}{\mu}\right). 
$ 
Moreover, these families can be found in polynomial time {\it poly}$(\ell,u)$.
\end{thm}

The results of Theorem~\ref{th.06} will be avoided in our further investigation because a presentation of the output sequence length $u$ in the form $\ell-\Delta$ is inconvenient in the optimization procedure.

We will get an estimate of $u$ taken from the results of Theorems~\ref{th.07}-\ref{th.09} directly. More specifically, using~(\ref{eq.035}) and~(\ref{eq.037}) one can write the relation for the necessary number of seed symbols for the first extractor~(\ref{eq.028}) in Theorem~\ref{th.06}:
\begin{equation}
u = \left\lceil\frac{\left\lceil\log\frac{k}{\epsilon}\right\rceil}{\ln c}\right\rceil\cdot\left\lceil\log\frac{k}{\epsilon}\right\rceil. \label{eq.039}
\end{equation}
For the second extractor~(\ref{eq.029}), it follows from the proof of lemma 17 in~\cite{xxxvi} that $u_0 = \left\lceil\frac{\nu}{\ln 2}\right\rceil\cdot\nu$, where $u=\tau\cdot u_0$, $\tau= \left\lceil\log\frac{4}{\mu}\right\rceil$, $0<\mu<\frac{1}{2}$. Then in terms of~(\ref{eq.037}) we get 
$$
u = \left\lceil\frac{\left\lceil\log\frac{k}{\epsilon}\right\rceil}{\ln 2}\right\rceil\cdot\left\lceil\log\frac{k}{\epsilon}\right\rceil\cdot\left\lceil\log\frac{4}{\mu}\right\rceil. 
$$

\subsection{Privacy amplification} \label{ssc.gee}

The procedure of privacy amplification (PA) at the final stage of the key generation between users A and B has been investigated in detail in~\cite{xiv,xxi}. PA can be implemented either by hashing~\cite{xiv,xxi} or by extraction~\cite{xxxv}. We will consider in the current paper the second approach. 

In order to compare our new results with the results obtained in~\cite{xxvi} where hashing has been used as the privacy amplification procedure, let us specify an application of both methods.

The sequence $X^k$ of length $k$ bits is mapped by the user A to the sequence $X_A^{\ell}$ of length $\ell$ through a keyless hash function from the class $U_2$, or $SU_2$. In a similar manner the user B forms his key $X_B^{\ell}$ after error-correcting his sequence $Y^k$.

The most important parameter of the PA procedure is the residual Shannon's information received by the adversary E, concerning the final key $K_A=K_B$. The estimates of this information leaking of E are given in~\cite{xiv,xxi} for different settings of wire-tap channels. The more general bound is presented below.

\begin{thm}[see~\cite{xiv}]\label{th.11} Let $X^k$ be the sequence of length $k$ transmitted from A to B over a BSC with the error probability $p_m$ and received by B as the sequence $Y^k$. Assume also that $Z^k$ is the result of receiving $X^k$ by the adversary E over a BSC with probability $p_w$, while the Renyi information contained in $Z^k$ about $X^k$ is $t$. Let $H$ be a $U_2$-class of hash functions from $\{0,1\}^k$ into $\{0,1\}^{\ell}$, known by all participants (A, B and E) and let $h\in H$ be a truly random hash function chosen by A, transmitted to B over a PDC. If the users A and B compute their keys as $K_A=h(X^k)$, $K_B=h(\tilde{Y}^k)$, then the amount of Shannon's information about the keys $K_A,K_B$ is upper bounded as
\begin{equation}
I(K_A;Z^k,h) \leq \frac{2^{-(k-\ell-t)}}{\ln 2}. \label{eq.041}
\end{equation}
\end{thm}

The Renyi information $t$ is connected with the Renyi entropy $H_2(X^k|Z^k)$ as
\begin{equation}
t = k - H_2(X^k|Z^k). \label{eq.042}
\end{equation}
For the BSC used as wire-tap channel we have by~(\ref{eq.021})
$$
H_2(X^k|Z^k) = -k\,\log\left(p_w^2 + (1-p_w)^2\right). 
$$
If the adversary receives some extra information about $X^k$ aside the information contained in $Z^k$ (for instance the sequence $C^r$ of check symbols of length $r$ eavesdropped by E over the PDC), then the amount of conditional Renyi entropy $H_2(X^k|Z^k, h,C^r)$ that received E can be estimated by the theorem~\ref{th.12} appearing below. (This theorem also gives the estimation of the amount of conditional minimal entropy $H_{\infty}(X^k|Z^k, h,C^r)$ which we use later).

\begin{thm}[see~\cite{xxi}]\label{th.12} Let $X$ and $C$ be two random variables and let $s>0$. Then the following inequalities hold:
\begin{equation}
H_2(X) - H_2(X|C=c) \leq \log|C| + s \label{eq.044}
\end{equation}
with a probability at least $1-2^{-\frac{s}{2}-1}$,
\begin{equation}
H_{\infty}(X^k) - H_{\infty}(X^k|C=c) \leq \log|C| + s \label{eq.045}
\end{equation}
with a probability at least $1-2^{-s}$.
\end{thm}

We can apply~(\ref{eq.044}) in order to estimate $H_2(X^k|Z^k, h,C^r)$. Then we get
\begin{equation}
H_2(X^k|Z^k, h,C^r) \geq H_2(X^k|Z^k,h) - r - s. \label{eq.046}
\end{equation}
It follows from~(\ref{eq.042}) and~(\ref{eq.046}) that
\begin{eqnarray}
\tilde{t} &=& k - H_2(X^k|Z^k, h,C^r) \nonumber \\
 &\leq& k - H_2(X^k|Z^k,h) + r + s \nonumber \\
 &=& t + r + s. \label{eq.047}
\end{eqnarray}
Substituting $\tilde{t}$ from~(\ref{eq.047}) into~(\ref{eq.041}) instead of $t$, we get the upper bound of Shannon's information leaking to E
$$
I(K_A;Z^k, h,C^r) \leq \frac{2^{-(k-\ell-t-r-s)}}{\ln 2}, 
$$
that holds with a probability
$
P_{risk} \leq 1- 2^{\frac{s}{2}-1}. 
$ 
In order to compare the performance of privacy amplification based on hashing and on extraction, let us prove a new lemma establishing a connection between the Shannon's information leaking to adversary regarding the key at the output of the extractor, and the statistical difference among distribution of the final key and an uniform distribution.

\begin{lemma}\label{lm.01} If the statistical distance between the output of the extractor generating the length $\ell$ key and an uniform distribution is at most $\epsilon$, then the amount of the Shannon's information concerning the key got by any adversary is upperly bounded as 
$$
I(K^{\ell};Z^k| \Gamma^u) \leq 2\ell\sqrt{\epsilon}. 
$$
\end{lemma}

{\em Proof.} The following inequality holds by definition of strong extractor
\begin{equation}
\dff(\Gamma^u\circ \mbox{\rm Ext}(Z^k,\Gamma^u),V^{\ell+u}) \leq \epsilon. \label{eq.051}
\end{equation}
The meaning of this inequality is that the probability distribution of the concatenation of the extractor's output and the ``seed'' $\Gamma^u$ is close enough to an uniform distribution. In order to simplify the notation, let us denote by $D$ the term at the left of~(\ref{eq.051}).
We note initially that for an uniform distribution on the space $V^{\ell+u}$, the following equality holds
\begin{equation}
D = M_{\Gamma^u}[\dff(\mbox{\rm Ext}(Z^k,\gamma^u),V^{\ell})]. \label{eq.052}
\end{equation}
where $\gamma^u$ is a random sequence 
and $M_{\Gamma^u}[\cdot]$ is the expectation with respect to the distribution on $\Gamma^u$.
In fact, $D$ itself can also be expressed as the term: 
\begin{equation}
\frac{1}{2}\sum_{\gamma^u,e^{\ell}\in E}\left| P_{E|\Gamma^u}(e^{\ell})\, P_{\Gamma^u}(\gamma^u) - P_{V^{\ell}}(v^{\ell}) P_{V^{u}}(v^{u})\right|, \label{eq.053}
\end{equation}
where $e^{\ell}$ is the output sequence of the extractor $E$.

Since the distributions $P_{\Gamma^u}(\gamma^u)$ and $P_{V^{u}}(v^{u})$ are both uniform, the term~(\ref{eq.053}) gives:
\begin{eqnarray*}
D &=& \frac{1}{2}\sum_{\gamma^u}P_{\Gamma^u}(\gamma^u) \sum_{e^{\ell}\in E}\left|P_{E|\Gamma^u}(e^{\ell}) - P_{V^{\ell}}(v^{\ell})\right| \\
 &=& \frac{1}{2}\sum_{\gamma^u}P_{\Gamma^u}(\gamma^u) \dff(\mbox{\rm Ext}(Z^k,\Gamma^u),V^{\ell}) \\
 &=& M_{\Gamma^u}[\dff(\mbox{\rm Ext}(Z^k,\gamma^u),V^{\ell})] 
\end{eqnarray*}
proving~(\ref{eq.052}). By combining~(\ref{eq.051}) and~(\ref{eq.052}) it is obtained
\begin{equation}
M_{\Gamma^u}[\dff(\mbox{\rm Ext}(Z^k,\gamma^u),V^{\ell})] \leq \epsilon. \label{eq.054}
\end{equation}
Using the well known Markov's inequality,~(\ref{eq.054}) implies
$$
P_{\Gamma^u}\left[\dff(\mbox{\rm Ext}(Z^k,\gamma^u),V^{\ell}) \leq \rho\epsilon\right] \geq 1 - \frac{1}{\rho}, 
$$
where $\rho \geq 1$ is some arbitrary value.

In the lemma 6 at~\cite{xxi}, the following inequality has been proved which put in our own  notation states 
\begin{equation}
H\left(\mbox{\rm Ext}(Z^k,\gamma^u)\right) \geq \ell \left(1 -\dff(\mbox{\rm Ext}(Z^k,\gamma^u),V^{\ell}) - 2^{-\ell} \right). \label{eq.056}
\end{equation}
Taking into account that the output extractor sequence is just the key, we can write
\begin{equation}
H\left(\mbox{\rm Ext}(Z^k,\gamma^u)\right) = H\left(K^{\ell}|\Gamma^u = \gamma^u\right). \label{eq.057}
\end{equation}
Then by substituting~(\ref{eq.057}) into~(\ref{eq.056}), we have that the inequality
\begin{equation}
H\left(K^{\ell}|\Gamma^u = \gamma^u\right) \geq \ell \left(1 -\dff(\mbox{\rm Ext}(Z^k,\gamma^u),V^{\ell}) - 2^{-\ell} \right) \label{eq.058a}
\end{equation}
will hold with probability
\begin{equation}
P_{\Gamma^u}\left[\mbox{inequality~(\ref{eq.058a}) holds}\right] \geq 1 - \frac{1}{\rho}. \label{eq.058}
\end{equation}
It follows from~(\ref{eq.058}) a trivial estimate for the averaged value $H\left(K^{\ell}|\Gamma^u = \gamma^u\right)$ over $\gamma^u$, namely
\begin{equation}
H\left(K^{\ell}|\Gamma^u\right) \geq \left(1 - \frac{1}{\rho}\right) \ell \left(1 - \rho\epsilon - 2^{-\ell}\right). \label{eq.059}
\end{equation}
After a simplification on the right side of~(\ref{eq.059}) and by neglecting smaller values than $2^{-\ell}$
$$
H\left(K^{\ell}|\Gamma^u\right) \geq \ell \left(1 - \rho\epsilon - \frac{1}{\rho} + \epsilon\right). 
$$
Then for the amount of information leaking of an adversary concerning the key $K^{\ell}$, given the knowledge of $\Gamma^u$, the following bound is obtained
\begin{equation}
I(K^{\ell};Z^k| \Gamma^u) \leq \frac{\ell}{\rho} + \ell\epsilon(\rho-1). \label{eq.061}
\end{equation}
The right side of~(\ref{eq.061}) is minimized under the condition $\rho = \frac{1}{\sqrt{\epsilon}}$ giving the final inequality
$$I(K^{\ell};Z^k| \Gamma^u) \leq 2\ell\sqrt{\epsilon}$$
providing thus the desired result. \hfill$\Box$
\medskip

It follows from the above lemma that if the value of the statistical difference at the extractor output that forms the length $\ell$ key does not exceed $\epsilon$, then the amount of the residual information regarding the key obtained by the adversary does not exceed $2\ell\sqrt{\epsilon}$.

This means that a requirement, regarding the amount of Shannon's information on the key leaking to an adversary, of the form $I(K^{\ell};Z^k| \Gamma^u) \leq I^{adm}$ will be fulfilled if $2\ell\sqrt{\epsilon} = I^{adm}$. This fact results in the following requirement to the extractor's statistical distance:
\begin{equation}
\epsilon = \left(\frac{I^{adm}}{2\ell}\right)^2. \label{eq.062}
\end{equation}

\section{Key distribution protocols} \label{sc.03}

\subsection{Statement of the protocols}

Two key distribution protocols in presence of an active adversary have been proposed by Maurer and Wolf in~\cite{xxi}: the {\em UH-protocol}, in which privacy amplification procedure was executed using hash functions and the {\em EX- protocol} based on extractions. It has been shown in~\cite{xxi} that the EX- protocol majors the UH-protocol with respect to several conditions.

We want to investigate a performance of these and other new protocols. We will show that our new protocols are superior than those considered in~\cite{xxi} for non-asymptotic cases (e.g. when the sequence lengths are finite).

Initially we consider modified UH- and EX-protocols and denote them as $\alpha$ and $\alpha_{ext}$, respectively. A difference between the original and the modified protocols is determined by two factors.
\begin{enumerate}
\item We consider protocols under the condition $\pi_A\not=0$, $\pi_B\not=0$, $\pi_A,\pi_B<\pi_E$, or equivalently the conditions $p_m>0$, $p_w>p_m$, see Figure~\ref{fig.01}. This requires to send the check symbols from A to B in order to conciliate $X^k$ and $Y^k$.
\item Instead of the authentication algorithm ``{\em request-response}'' presented in~\cite{xxi}, we will use non-interactive the AC-based algorithm (see Section~\ref{sc.02}) because this allows the users to provide authentication even when the sequences $X^k$ and $Y^k$ do not coincide completely. By the same reason, the authentication algorithm and the number of substrings of the original strings $X^k$ and $Y^k$ are changed. 
\end{enumerate}
Before the execution of the $\alpha$, $\alpha_{ext}$-protocols, the users A and B divide their respective sequences $X^k$, $Y^k$, into $X_1^{k_1}$, $X_2^{k_2}$ and $Y_1^{k_1}$, $Y_2^{k_2}$ of lengths $k_1, k_2$. (The first parts $X_1^{k_1}$ and $Y_1^{k_1}$ will be used for key generation in the execution of the PA procedure while the second parts $X_2^{k_2}$ and $Y_2^{k_2}$ will be used in the execution of the authentication procedure).
Since the $\alpha$-protocol was already considered in~\cite{xxvi}, we move on to the $\alpha_{ext}$-protocol~\cite{xxix}.
\begin{enumerate}
\item The user A forms the string $C_1^{r_1}$ of check symbols of length $r_1$ to the string $X_1^{k_1}$ using a $(k_1+r_1, k_1)$-error correcting code ${\cal C}_1$. (This code should be agreed by users in advance.)
\item The user A generates a truly random binary sequence $\gamma$ (which will be used as an extractor seed) of length $u$. 
\item The user A forms the authenticator ${\bf w}$ for the message $(C_1^{r_1},\gamma)$ using for that an AC based on an error correcting $(n_0, k_0 = r_1+u, d)$-code and the sequence $X_2^{k_2}$.
\item The user A sends to B the message $(C_1^{r_1},\gamma)$ over a PDC appended with the authenticator ${\bf w}$.
\item The user B verifies the authenticity of the message $(C_1^{r_1},\gamma)$ through the known $(n_0, k_0)$-AC and his string $Y_2^{k_2}$ (see section~\ref{ssc.auth}). If authenticity is confirmed, then B goes to the next step. Otherwise he rejects the KDP.
\item The user B corrects the error in string $Y_1^{k_1}$ through the check symbols string $C_1^{r_1}$. We denote by $\tilde{Y}_1^{k_1}$ the string $Y_1^{k_1}$ after error correction.
\item In order to get the keys $K_A$ and $K_B$ both users A and B execute a privacy amplification procedure based on an extractor (see section~\ref{ssc.ext}): $K_A = E_{ext}(X_1,\gamma)$, $K_B = E_{ext}(\tilde{Y}_1,\gamma)$.
\end{enumerate}
Recall that the $\alpha$-protocol differs from the $\alpha_{ext}$-protocol in that it generates a hash function $h$ in step 2. This hash function jointly with the check symbols of $C_1^{r_1}$ and the authenticator {\bf w} are transmitted to B (steps 3-5). In the seventh step, this hash function is needed for key generation: $K_A = h(X_1)$, $K_B = h(\tilde{Y}_1)$.

It has also been proposed in~\cite{xxvi} a new $\beta$-protocol that differs from the $\alpha$-protocol in the following: After the execution of the initialization phase, both users A and B have got the strings that can in fact be used to form the hash functions. In this way, we do not require to send the hash functions over the PDC, hence the length $k_2$ used before for authentication of the hash function can be shortened. Therefore we may expect that the length of the substring $X_1$ is increased (if the total length of the string $X$ is fixed). But such conclusion is not so apparent because we have to extract the hash function as a segment from the string $X$.

A similar problem appears in the case in which an extractor is used instead of a hash function for privacy amplification. In the $\alpha_{ext}$-protocol A generates a truly random sequence $\gamma$ and sends it to B jointly with the authenticator of $\gamma$. But the required sequence $\gamma$ can be gotten directly by both users A and B from the initially distributed strings $X$ and $Y$. This results in the following $\beta_{ext}$-protocol.
It is worth to note that although $\gamma$ is not uniformly distributed from the adversary's point of view this has no relevance for strong extractors. 

Within the above setup, the users A and B divide the strings $X^k$, $Y^k$ into three disjoint parts $X_1^{k_1}$, $X_2^{k_2}$, $X_3^{k_3}$ and $Y_1^{k_1}$, $Y_2^{k_2}$, $Y_3^{k_3}$ with $k_1 + k_2 + k_3 = k$. Then they execute the following steps:
\begin{enumerate}
\item The user A forms the length $r_1$ string $C_1^{r_1}$ of check symbols of the string $X_1^{k_1}$ using the error correcting $(k_1+r_1,k_1)$-code ${\cal C}_1$, agreed in advance.
\item The user A forms the length $r_2$ check string $C_2^{r_2}$ of the string $X_3^{k_3}$ using the error correcting $(k_3+r_2,k_3)$-code ${\cal C}_2$, agreed in advance.
\item The user A forms the authenticator ${\bf w}$ of the message $(C_1^{r_1},C_2^{r_2})$ using an AC and his substring $X_2^{k_2}$.
\item The user A sends to B the message $(C_1^{r_1},C_2^{r_2})$ over a PDC appended with ${\bf w}$.
\item The user B verifies the authenticity of the message $(C_1^{r_1},C_2^{r_2})$ using a AC and his substring $Y_2^{k_2}$. If it is confirmed then he goes to the next step. Otherwise he rejects the KDP.
\item The user B corrects errors on strings $Y_1^{k_1}$, $Y_3^{k_3}$, using the check strings $C_1^{r_1}$ and $C_2^{r_2}$. Denote by $\tilde{Y}_1^{k_1}$, $\tilde{Y}_3^{k_3}$, the strings $Y_1^{k_1}$, $Y_3^{k_3}$, after error corrections.
\item The users A and B take their substrings $X_3^{k_3}$, $\tilde{Y}_3^{k_3}$, where $k_3=u$, as the second argument $\gamma^u$ in their extractors.
\item Both users A and B form the keys as $K_A = E_{ext}(X_1,X_3)$, $K_B = E_{ext}(\tilde{Y}_1,\tilde{Y}_3)$.
\end{enumerate}

\subsection{Performance evaluation of the protocols}

A theorem has been proved in~\cite{xxvi} determining the optimal parameters for both the $\alpha$, and $\beta$-protocols depending on the posed requirements. Let us prove a generalization of that theorem for the $\alpha$, $\beta$, $\alpha_{ext}$, and $\beta_{ext}$-protocols. We will assume that for the $\alpha$, and $\beta$-protocols a hashing is used as privacy amplification procedure, whereas for $\alpha_{ext}$, and $\beta_{ext}$-protocols an extraction is used. Moreover we assume that the first extraction scheme considered in section~\ref{ssc.ext} is used, where the number of random bits u is determined by equation~(\ref{eq.039}).

\begin{thm}\label{th.13}
Let us assume that the users A, B and the adversary E have binary strings $X^k$, $Y^k$ and $Z^k$, respectively after execution of the initialization phase over the wire-tape channel, $p_w=\pr{x_i\not=y_i}$, $p_w=\min\{\pr{x_i\not=z_i},\pr{y_i\not=z_i}\}$, $p_m\geq 0$, $p_w>p_m$. Then A and B are able to form a common key of length $\ell$ satisfying the requirements~(\ref{eq.003})-(\ref{eq.007}) after the execution of any of the $\alpha$, $\beta$, $\alpha_{ext}$, and $\beta_{ext}$-protocols if the parts of lengths $k_1,k_2$ on which were divided the substrings $X^k$, $Y^k$ for the $\alpha$, and $\alpha_{ext}$-protocols or the parts of lengths $k_1,k_2,k_3$ on which were divided the substrings $X^k$, $Y^k$ for the $\beta$, and $\beta_{ext}$-protocols satisfy the equations listed below: 
\begin{itemize}
\item for all protocols
\begin{equation}
k_1 = -\frac{\log P^{adm}_e}{E(R_{c_1})}, \label{eq.063}
\end{equation}
\item for $\alpha$ and $\beta$-protocols
\begin{equation}
k_1 = \frac
{\ell + r_1 -2\log P^{adm}_{risk} - \log(I^{adm}\ln 2) - 2}
{-\log(p_w^2+(1-p_w)^2)}, \label{eq.064}
\end{equation}
\item for $\alpha_{ext}$, and $\beta_{ext}$-protocols
\begin{eqnarray}
k_1\cdot H_{\infty} &=& \ell c + r_1 - \log P^{adm}_{risk} + u \nonumber \\
 & & + 3 \log\ell\ \left(\frac{I^{adm}}{2\ell}\right)^{-2} + 3
, \label{eq.065}
\end{eqnarray}
\end{itemize}
where $R_{c_1}$ and $E(R_{c_1})$ are determined by~(\ref{eq.011})-(\ref{eq.014}) and 
\begin{equation}
u = \left\lceil\frac{\left\lceil\log k_1\,\left(\frac{I^{adm}}{2\ell}\right)^{-2}\right\rceil}{\ln c}\right\rceil\cdot\left\lceil\log k_1\,\left(\frac{I^{adm}}{2\ell}\right)^{-2}\right\rceil \label{eq.066}
\end{equation}
is the number of the extractor random symbols, $c$ is a parameter under optimization, $H_{\infty} = -\log\max(p_w,1-p_w)$,
\begin{eqnarray}
k_2\left(1-g\left(\frac{2d}{k_2}\right)\right) &=& 2k_0, \label{eq.067} \\
\sum_{i=\Delta_w+1}^{k_2} {k_2\choose i} p_m^i(1-p_m)^{k_2-i} &=& P_f^{adm}, \label{eq.068} \vspace{3ex}\\
\sum_{i=0}^{\Delta_w} {d\choose i} p_w^i(1-p_w)^{d-i}\cdot\hspace{6em} & & \nonumber \\
\sum_{j=0}^{\Delta_w-i} {k_2-d\choose j} p_m^j(1-p_m)^{k_2-d-j} &=& P_d^{adm}, \label{eq.069} 
\end{eqnarray}
where
\begin{equation}
k_0 = \left\{\begin{array}{ll}
k_1+r_1 & \mbox{for the $\alpha$-protocol,} \\
2r_1 & \mbox{for the $\beta$-protocol,} \\
u+r_1 & \mbox{for the $\alpha_{ext}$-protocol,} \\
r_1+r_2 & \mbox{for the $\beta_{ext}$-protocol,} 
\end{array}\right. \label{eq.070}
\end{equation}
and $r_2$ being the number of check symbols of the error correcting $(k_3+r_2,k_3)$-code ${\cal C}_2$ found similarly as in equation~(\ref{eq.063}), 
\begin{equation}
k_3 = \left\{\begin{array}{ll}
0 & \mbox{for the $\alpha$-protocol,} \\
k_1 & \mbox{for the $\beta$-protocol,} \\
0 & \mbox{for the $\alpha_{ext}$-protocol,} \\
u & \mbox{for the $\beta_{ext}$-protocol.} 
\end{array}\right. \label{eq.071}
\end{equation}
The key rate is then determined as follows: 
\begin{eqnarray}
R_{\alpha} = \frac{\ell}{k_1+k_3} &,& R_{\beta} = \frac{\ell}{2k_1+k_3}, \nonumber \\ 
R_{\alpha_{ext}} = \max_c\frac{\ell}{k_1+k_2} &,& R_{\beta_{ext}} = \max_c\frac{\ell}{u+k_1+k_2}. \label{eq.073} 
\end{eqnarray}
\end{thm}

{\em Proof}.
For the $\alpha$ and $\beta$-protocols the theorem has been proved in~\cite{xxvi}. Let us prove it only for the $\alpha_{ext}$, and $\beta_{ext}$-protocols.

Let the bounds of the KDP parameters meet exactly all requirements~(\ref{eq.003})-(\ref{eq.007}), e.g. the following equation hold:
$$
P_e = P_e^{adm} = 2^{-k_1E(R_{c1})} 
$$
where $R_{c1}=\frac{k_1}{k_1+r_1}$ is the code rate, and $E(R_{c1})$ is computed by~(\ref{eq.011})-(\ref{eq.013}). Under the condition that the adversary gets the sequence $Z^{k_1}$ over a BSC with error probability $p_w$ the conditional minimal entropy is
\begin{eqnarray*}
H_{\infty}\left(X^{k_1}|Z^{k_1}\right) &=& k_1H_{\infty}\left(X|Z\right) \\
 &=& -k_1\log\max(p_w,1-p_w) \\
 &=& k_1H_{\infty}.
\end{eqnarray*}
Since the adversary receives also the check block $C_1^{r_1}$, in line with~(\ref{eq.045}) the following inequality results:
\begin{equation}
H_{\infty}\left(X^{k_1}|Z^{k_1},C_1^{r_1}\right) \geq k_1H_{\infty} - r_1 - s, \label{eq.075}
\end{equation}
which does not comply with the probability $P_{risk} \leq 2^{-s}$. 

By substituting~(\ref{eq.075}) into~(\ref{eq.036}), we may write $\varepsilon\leq 2^{\frac{\tau}{3}}$ where $\tau = \ell c + 3\log\ell - k_1H_{\infty}+r_1+s+u+3$. Let us assume that $I^{adm}$ is chosen in such a way that 
$$2^{\frac{\tau}{3}} = \log\left(\frac{I^{adm}}{2\ell}\right)^2,$$
resulting thus condition~(\ref{eq.062}). Hence we can write 
\begin{equation}
\ell c + 3\log\ell - k_1H_{\infty} + r_1+ s + u + 3 = 3\log\left(\frac{I^{adm}}{2\ell}\right)^2. \label{eq.076}
\end{equation}
Assuming $P_{risk} =  P_{risk}^{adm} = 2^{-s}$,~(\ref{eq.076}) holds eventually from~(\ref{eq.065}). 
The value $u$ in~(\ref{eq.076}) is the number of the extractor random symbols. In order to find it, we can use~(\ref{eq.039}) substituting $\epsilon$ by $\left(\frac{I^{adm}}{2\ell}\right)^2$ in line with Lemma~\ref{lm.01} that results in~(\ref{eq.066}). A solution of the equation system~(\ref{eq.063}) and~(\ref{eq.065}) allows to find the parameters $k_1$, $r_1$, given a fixed $c$. It will be shown in the sequel that the key rate can be maximized by a proper selection of the parameter $c$.

In order to find $k_2$ let us assume that the probabilities $P_f$ and $P_d$ have equal values, $P_f = P_f^{adm}$, $P_d = P_d^{adm}$, with
\begin{eqnarray}
P_f^{adm} &=& \sum_{i=\Delta_w+1}^{2n_0} {2n_0 \choose i} p_m^i (1-p_m)^{2n_0-i}, \label{eq.077} \\
P_d^{adm} &=& \sum_{i=0}^{\Delta_w} {d \choose i} p_w^i (1-p_w)^{d-i}\cdot \nonumber \\
 & & \ \sum_{j=0}^{\Delta_w-i} {2n_0-d \choose j} p_m^j (1-p_m)^{2n_0-d-j}, \label{eq.078} 
\end{eqnarray}
where $n_0,k_0,d$ are the parameters of error correcting codes used in the AC.

Recall that for the AC we had $k_2 = 2n_0$ where $n_0$ is the length of the error correcting $(n_0,k_0)$-code with minimum distance $d$. For the $\alpha_{ext}$-protocol $k_0 = r_1 + u$, while for the $\beta_{ext}$-protocol $k_0 = r_1 + r_2$, where $r_2$ is the number of check symbols in the $(k_3+r_2,k_3)$-code ${\cal C}_2$. This gives relation~(\ref{eq.070}) for the parameter $k_0$.

Using the Varshamov-Gilbert inequality~\cite{xxxix} connecting $n_0,k_0,d$ and taking into account that $k_2 = 2n_0$ we get
\begin{equation}
k_2\left(1-g\left(\frac{2d}{k_2}\right)\right) = 2 k_0. \label{eq.079}
\end{equation}
Solving the equation system~(\ref{eq.077})-(\ref{eq.079}), equivalent to the equation system~(\ref{eq.067})-(\ref{eq.069}), we find the parameters $k_2, d$. The value $r_2$ is calculated by~(\ref{eq.011})-(\ref{eq.013}), in which it is necessary to let $k=ur=r_2$, $p=p_m$, $P_e = P_e^{adm}$.

In line with the above protocols, we have that for the $\alpha_{ext}$-protocol, $k_3=0$ and for the $\beta_{ext}$-protocol, $k_3=u$. This fact proves~(\ref{eq.071}). Then relation~(\ref{eq.073}) is apparent from the protocols description.

\begin{rmk}\label{rm.01} If the solution of the system~(\ref{eq.063})-(\ref{eq.071}) is not unique then it is reasonable to select any of them maximizing the key rate.
\end{rmk}

\begin{rmk}\label{rm.02} It is worth to note that the values $k_1,k_2,k_3,c$ found for the same requirements $P_e^{adm}$, $I^{adm}$, $P_{risk}^{adm}$, $P_f^{adm}$, $P_d^{adm}$, but for different protocols, may be different.
\end{rmk}

\begin{figure}[!t]
\centering
\begin{tabular}{c}
 \includegraphics[width=3in]{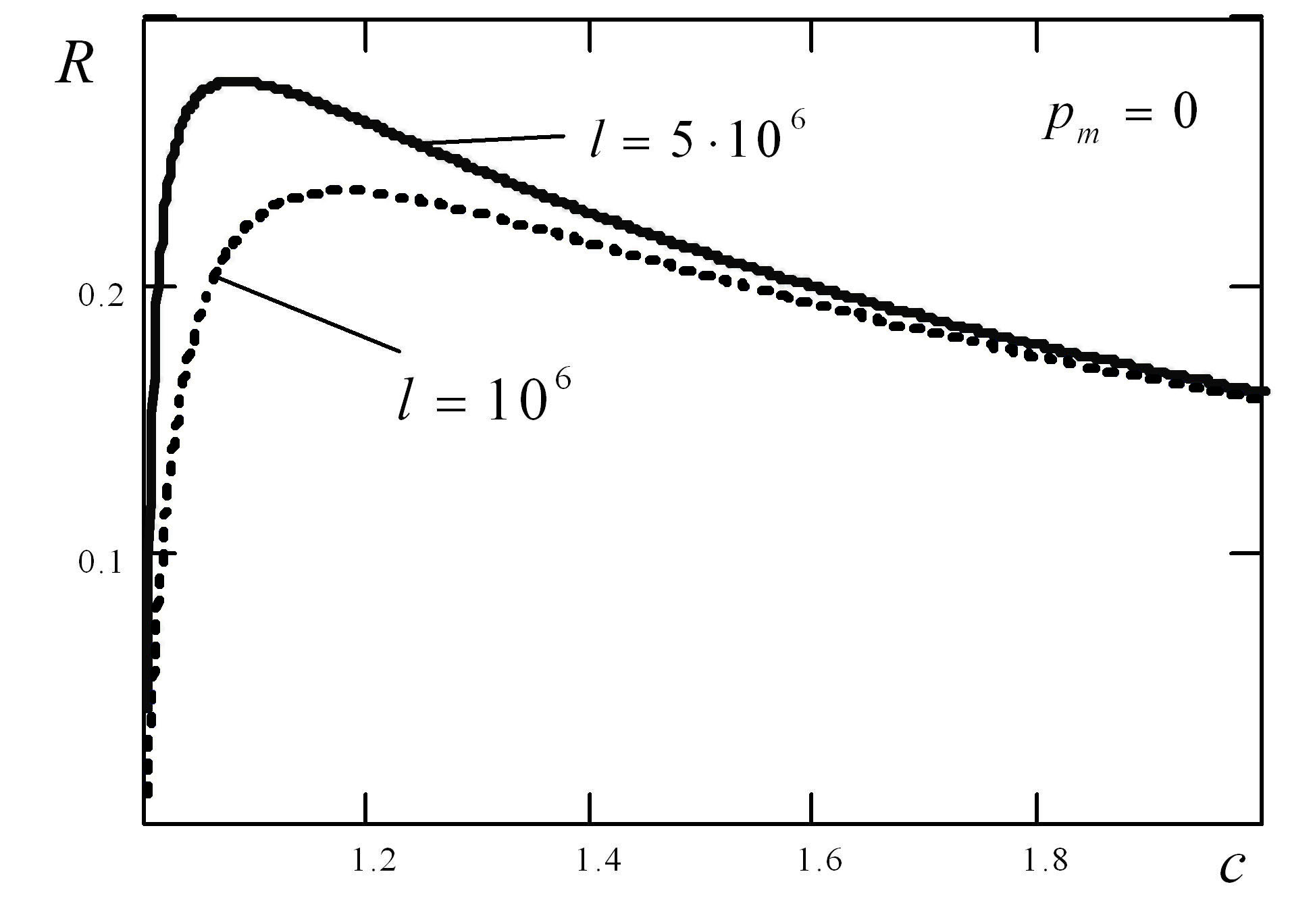} \hspace{1em} \\
 \includegraphics[width=3in]{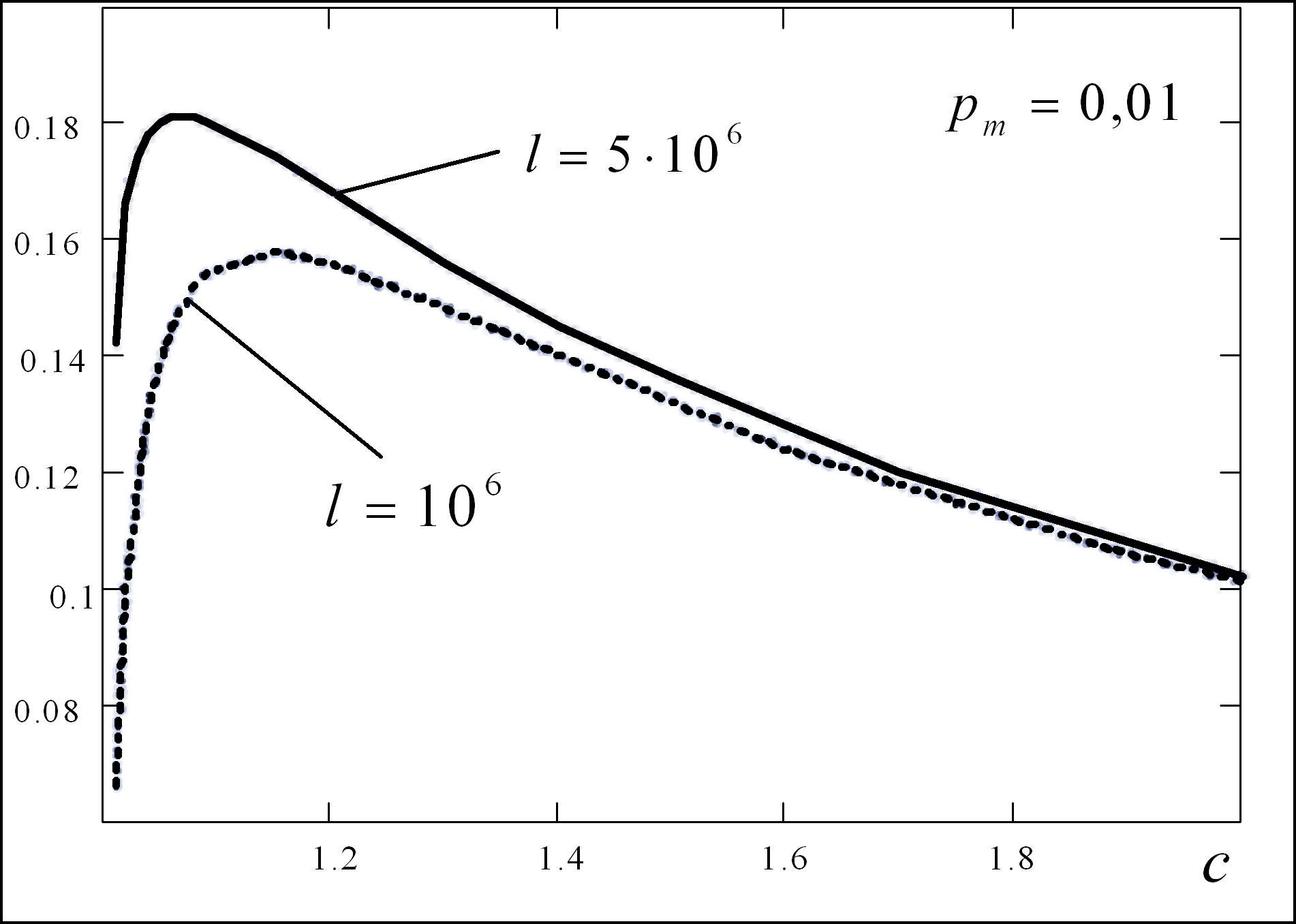}
\end{tabular}
\caption{Dependence of key rate versus the parameter extractor $c$ for the $\alpha_{ext}$-protocol.}
\label{fig.03}
\end{figure}

\begin{figure}[!t]
\centering
\begin{tabular}{c}
 \includegraphics[width=3in]{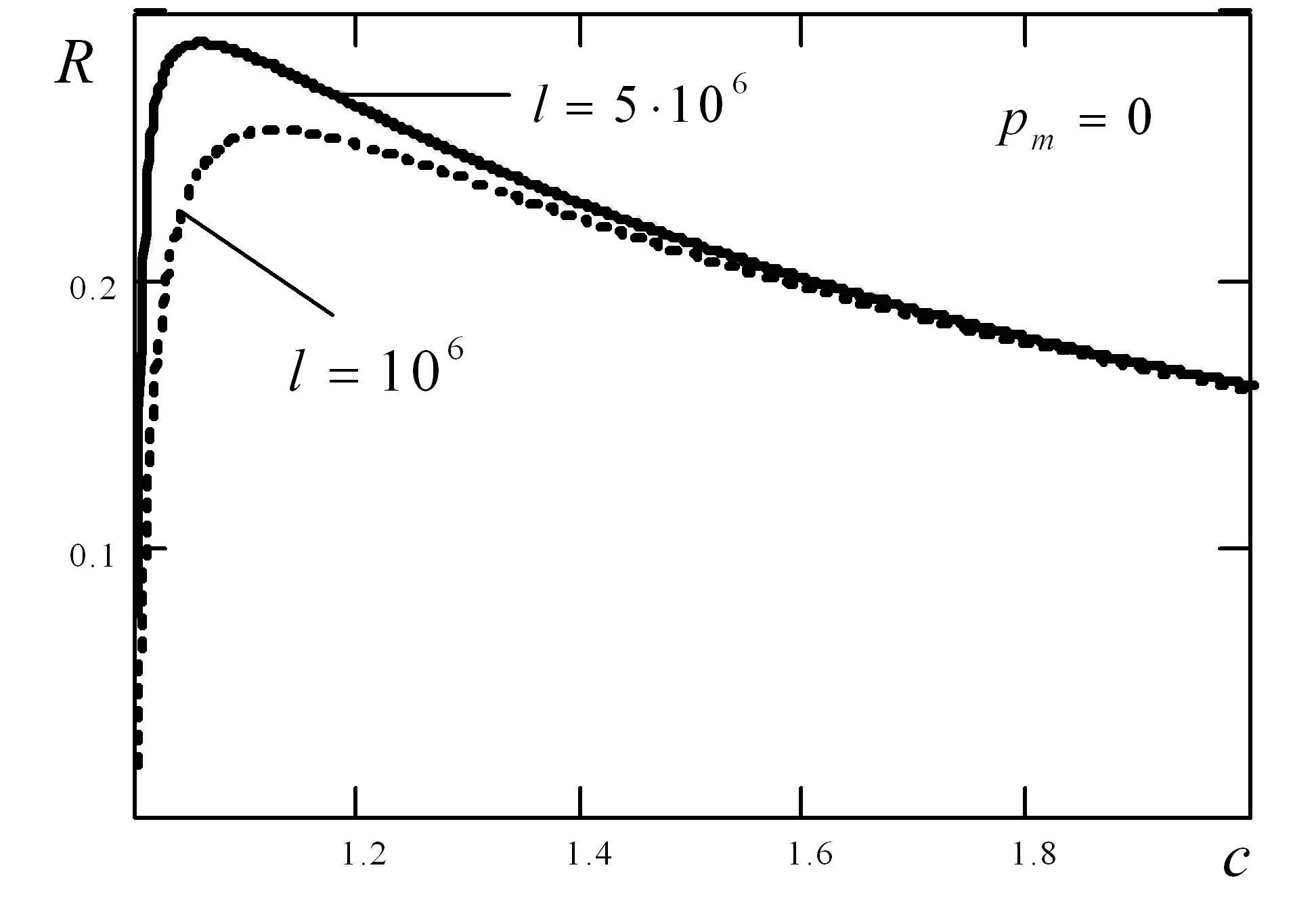} \\
 \includegraphics[width=3in]{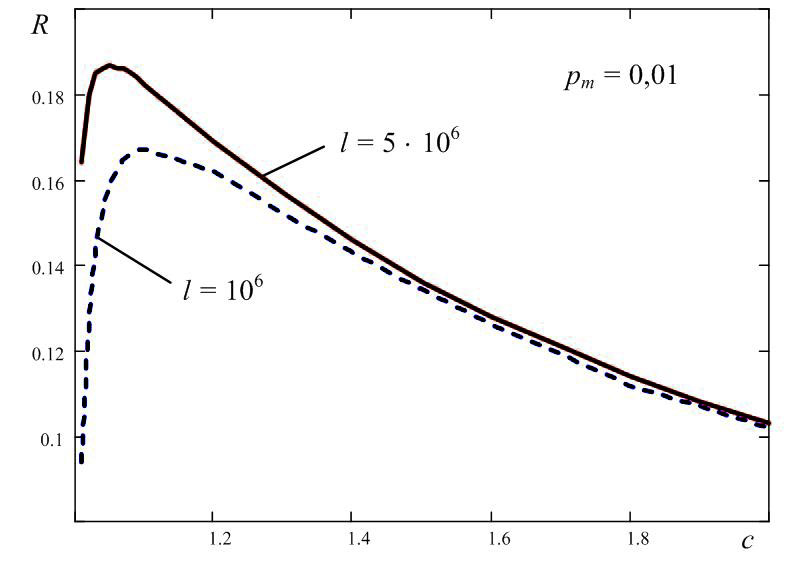} \, 
\end{tabular}
\caption{Dependence of key rate versus the parameter extractor $c$ for the $\beta_{ext}$-protocol.}
\label{fig.04}
\end{figure}

\begin{rmk}[Choice of $c$ in~(\ref{eq.065})-(\ref{eq.066})]\label{rm.03} In figures~\ref{fig.03} and~\ref{fig.04} the dependence of the key rate for the $\alpha_{ext}$-protocol and the $\beta_{ext}$-protocol is plotted versus the parameter $c$, given fixed values $\ell$ for different error probabilities in the main channels.
\end{rmk}

We assume that $p_w = 0.2$, $I^{adm}=10^{-30}$, $P_e^{adm} = P_{risk}^{adm} = P_f^{adm} = P_d^{adm} = 10^{-5}$ in the plotting of these curves. From these curves it is patent that the key rate depends essentially on the choice of the parameter $c$.

Let us compare the $\alpha$, $\beta$, $\alpha_{ext}$, $\beta_{ext}$-protocols on the key rates.

\begin{thm}\label{th.10} If the key length $\ell$ is given and the rate $R_c$ of the error correction code satisfies $R_c\geq\frac{2}{3}$, then $R_{\beta}\geq R_{\alpha}$, and $R_{\beta_{ext}}\geq R_{\alpha_{ext}}$.
\end{thm}

{\em Proof}. The first inequality is proved as theorem 10 in~\cite{xxvi}. Let us prove the second inequality. Let us write $R_{\alpha_{ext}} = \frac{\ell}{k_1+k_2}$ and $R_{\beta_{ext}} = \frac{\ell'}{k_1'+u+k_2'}$. For a fixed common length, $\ell=\ell'$, we should prove $k_1'+u+k_2'\leq k_1+k_2$.

Under the requirements~(\ref{eq.002}),~(\ref{eq.003}),~(\ref{eq.007}) posed to the KDP parameters, $P_{risk}$, $I^{adm}$, it follows from~(\ref{eq.065}) that $\ell' = \ell$ whenever $k_1 = k'_1$ and $r_1 = r_1'$. Therefore it is necessary just to prove
\begin{equation}
k_2\geq u+k_2' . \label{eq.079a}
\end{equation}
According to the scheme of the AC code design we can write $k_2 = 2n_0 = 2(r_1+u+r_0)$, where $r_0$ is the number of check symbols of the $(n_0,k_0)$-code. $k_2' = 2n_0' = 2(r_1+r_2'+r_0')$, where $r'_2$ is the number of check symbols of the $(u+r'_2, r'_2)$-code ${\cal C}_2$, and $r'_0$ is the number of check symbols of the $(n'_0,k'_0)$-code. 

By substituting the expressions for $k_2$, $k'_2$ presented above into~(\ref{eq.079a}), we get the equivalent inequality $u-2 r'_2+2r_0-2r_0'\geq 0$. In order to prove this inequality it is sufficient to show that $u \geq 2r'_2$ and $r_0\geq r'_0$.

The first inequality holds because under the theorem's condition, $R_c \geq \frac{2}{3}$ for the $(u+r'_2, r'_2)$-code. In order to prove the second inequality, we note that $r_0$ is the number of check symbols of the information block of length $k_0 = u+r_1$ and $r'_0$ is the number of check symbols of the information block of length $k_0' = r_1+r_2'$. It is clear that $k_0\geq k_0'$ and it is followed from Varshamov-Gilbert inequality that $\frac{d}{n_0}$ maintains a constant value as the information block length increases in order to get the required error correction capability. Therefore $r_0\geq r_0'$ and this completes the proof of the theorem. \hfill $\Box$
\medskip

With the purpose of comparing the protocols performance with hashing and with extraction, let us find the relation of key rate for sufficiently large $\ell$.

\begin{thm}\label{th.14} As the key length $\ell\to\infty$, then the following relations hold
\begin{eqnarray}
R_{\alpha} &=& \frac{H_2(p_w) - g(p_m)}{3+2g(p_m)}, \label{eq.080} \\
R_{\beta} &=& \frac{H_2(p_w) - g(p_m)}{2+4g(p_m)}, \label{eq.081} \\
R_{\alpha_{ext}} = R_{\beta_{ext}} &=& \frac{H_{\infty}(p_w) - g(p_m)}{1+2g(p_m)} .\label{eq.082} 
\end{eqnarray}
\end{thm}

{\em Proof.}
The proofs of~(\ref{eq.080}) and~(\ref{eq.081}) were presented in~\cite{xxvi}. In order to prove~(\ref{eq.082}), let us write the relations of the key rates at the $\alpha_{ext}$ and $\beta_{ext}$-protocols taking into account~(\ref{eq.025}),~(\ref{eq.070}),~(\ref{eq.071}),~(\ref{eq.073}):
\begin{eqnarray}
R_{\alpha_{ext}} &=& \frac{\ell}{k_1+2n_0} \nonumber \\
 &=& \frac{\ell}{k_1+k_2} \nonumber \\
 &=& \frac{\ell}{k_1+2(k_0+r_0)} \nonumber \\
 &=& \frac{\ell}{k_1+2u+2r_1+2r_0}, \label{eq.083} 
\end{eqnarray}

\begin{eqnarray}
R_{\beta_{ext}} &=& \frac{\ell}{k_1+u+2n_0} \nonumber \\
 &=& \frac{\ell}{k_1+u+k_2} \nonumber \\
 &=& \frac{\ell}{k_1+u+2(k_0+r_0)} \nonumber \\
 &=& \frac{\ell}{k_1+u+2u+2r_1+2r_0}, \label{eq.084}
\end{eqnarray}
where $u$ is the length of the extractor seed, $r_1$ is the length of the check string for the $(k_1+r_1,k_1)$-code, $r_2$ is the length of the check symbols string for the $(u+r_2,u)$-code, and $r_0$ is the length of the check string for the AC-code.

According to~(\ref{eq.016}), $r = k\,g(p_m)$ for sufficiently large $\ell$ (and hence sufficiently large $k$). Let us rewrite~(\ref{eq.083}),~(\ref{eq.084}) as 
\begin{eqnarray}
R_{\alpha_{ext}} &=& \frac{\ell}{k_1(1+2g(p_m))+2u+2r_0}, \label{eq.085} \\
R_{\beta_{ext}} &=& \frac{\ell}{(k_1+u)(1+2g(p_m))+2r_0}. \label{eq.086}
\end{eqnarray}
According with~(\ref{eq.065}),
\begin{equation}
k_1 =\frac
{\ell c - 2\log P_{risk} + u + 3 \log\ell\ \left(\frac{I^{adm}}{2\ell}\right)^{-2} + 3}
{H_{\infty}-g(p_m)}. \label{eq.087}
\end{equation}
Substituting $k_1$ into~(\ref{eq.085}) produces 
\begin{equation}
R_{\alpha_{ext}} = \frac{H_{\infty}-g(p_m)}
{c(1+2g(p_m))+2\left(H_{\infty}-g(p_m)\right)\left(\frac{u}{\ell}+\frac{r_0}{\ell}\right)} .\label{eq.088}
\end{equation}
It is easy to show that 
$$
\lim_{\ell\to\infty}\frac{u}{\ell} = 0. 
$$
Also theorem~\ref{th.05} establishes that $\lim_{\ell\to\infty}\frac{r_0}{k_0+r_0} = 0$, but since $k_0\to\infty$ as long as $\ell\to\infty$, according to~(\ref{eq.065}) and~(\ref{eq.070}) we get $\lim_{\ell\to\infty}\frac{r_0}{\ell} = 0$. Now we can write~(\ref{eq.088}), in the limit $R_{\alpha_{ext}}=\frac{H_{\infty}(p_w)-g(p_m)}{c(1+2g(p_m))}$, which approaches to a maximum as $c\to 1$. This provides a proof of~(\ref{eq.082}) for the $\alpha_{ext}$-protocol.

Similarly, by expressing $R_{\beta_{ext}}$ as~(\ref{eq.086}) using $k_1$ as in~(\ref{eq.087}), the used arguments in the proof of~(\ref{eq.082}) for the $\alpha_{ext}$-protocol, show that~(\ref{eq.082}) holds also for the $\beta_{ext}$-protocol. \hfill $\Box$
\medskip

The following trivial corollary results from the above theorem.

\begin{corol}\label{co.01} If the channel parameters $p_m$ and $p_w$ are such that
\begin{equation}
\frac{H_{\infty}(p_w)-g(p_m)}{1+2g(p_m)} \geq \frac{H_2(p_w)-g(p_m)}{3+2g(p_m)} \label{eq.090}
\end{equation}
and
\begin{equation}
2\left(H_{\infty}(p_w)-g(p_m)\right) \geq H_2(p_w)-g(p_m), \label{eq.091}
\end{equation}
then $R_{\alpha_{ext}}\geq R_{\alpha}$ and $R_{\beta_{ext}}\geq R_{\beta}$ respectively for sufficiently large $\ell$.
\end{corol}

\begin{corol}\label{co.02} If $p_m=0$, then $R_{\alpha_{ext}}\geq R_{\alpha}$ and $R_{\beta_{ext}}\geq R_{\beta}$.
\end{corol}

{\em Proof.}
If $p_m=0$, then the relations~(\ref{eq.090}),~(\ref{eq.091}) can be written as 
$$
H_{\infty}(p_w) \geq \frac{1}{3}H_2(p_w) \ \ \mbox{ and }\ \ H_{\infty}(p_w) \geq \frac{1}{2}H_2(p_w). 
$$
Since $2H_{\infty}(p_w)\geq H_2(p_w)$~\cite{xviii}, then $R_{\alpha_{ext}}\geq R_{\alpha}$ and $R_{\beta_{ext}}\geq R_{\beta}$. \hfill $\Box$
\medskip

\begin{figure}[!t]
\centering
\begin{tabular}{c}
 \includegraphics[width=3in]{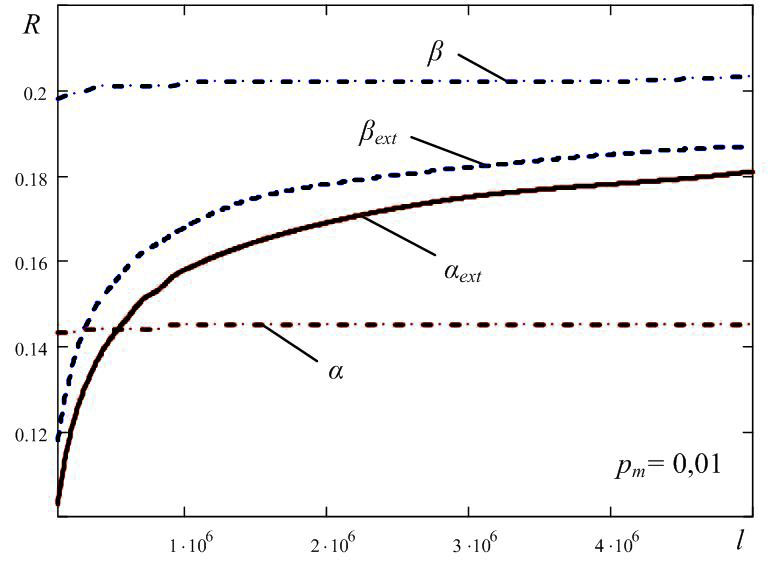} \\
 \includegraphics[width=3in]{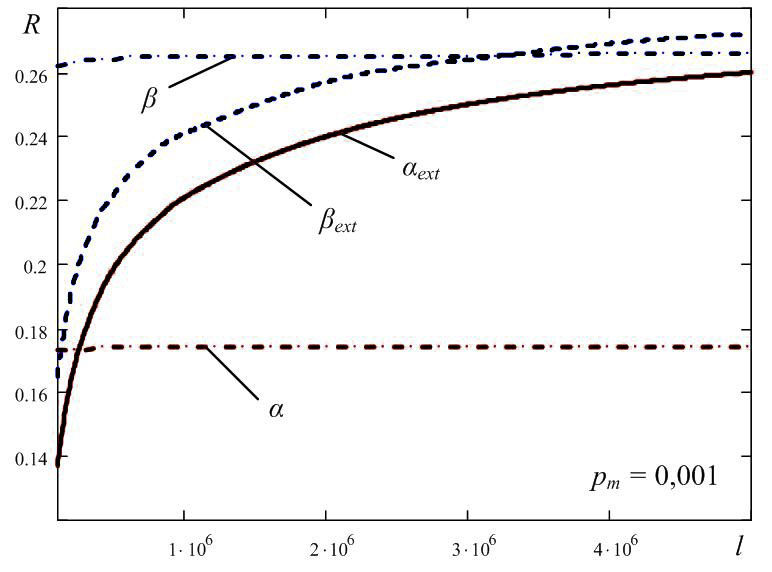} 
\end{tabular}
\caption{The key rates versus their lengths for different requirements imposed to KDPs.}
\label{fig.05}
\end{figure}

Let us exemplify the above results and illustrate that the $R_{\alpha_{ext}}(\ell)$ and $R_{\beta_{ext}}(\ell)$-protocols major the $R_{\alpha}(\ell)$ and $R_{\beta}(\ell)$-protocols respectively. Let us select the following natural requirements for the KDP: 
\begin{equation}
I^{adm} = 10^{-30}\ ,\ P_d = P_e = P_f = P_{risk} =10^{-5}. \label{eq.093}
\end{equation}
In figure~\ref{fig.05} we plot the key rates $R_k$ versus its length $\ell$ for both $R_{\alpha_{ext}}(\ell)$ and $R_{\beta_{ext}}(\ell)$-protocols with $p_m = 0.01$ and 0.001, $p_w = 0.2$, and the requirements presented in~(\ref{eq.093}).

The optimization of $c$ has been performed for every value of $\ell$. For comparison purposes the dependences $R_{\alpha}(\ell)$ and $R_{\beta}(\ell)$ are shown also in the figure. 

The following conclusions are drawn immediately after an examination of the obtained dependence.

The protocols using extractors have greater key rate than the $\alpha$ and $\beta$-protocols under sufficiently large $\ell$ and small $p_m$. It is worth to note that if for the $\alpha$ and $\beta$-protocols the asymptotically possible value key rate calculated by~(\ref{eq.080}),~(\ref{eq.081}) can be achieved even in the considered key length range, it is not true for the $\alpha_{ext}$ and $\beta_{ext}$-protocols, demonstrating a noticeable increasing proliferation of the key rate outside this range.

We can see that the $\alpha_{ext}$-protocol is superior than the $\alpha$-protocol when $\ell>5\cdot 10^5$ with $p_m=0.01$ and the $\beta_{ext}$-protocol is superior than the $\beta$-protocol when $\ell>3.5\cdot 10^6$ and $p_m=0.001$ (under the stated requirements in our investigations). The key length for which the $\alpha_{ext}$ and $\beta_{ext}$-protocols are superior than the $\alpha$ and $\beta$-protocols essentially depends on the error probabilities in the communication channels.

The $\beta_{ext}$-protocol is superior than the $\alpha_{ext}$-protocol, although these protocols have the same asymptotic key rate. Hence, protocols with extractors are superior than the protocols with hashing, when $p_m=0$.

\section{Key distribution protocols under the condition that legal users shared short authentication key before starting the KDP} \label{sc.04}

The $\alpha'$ and $\beta'$-protocols have been introduced in~\cite{xxvi}, which differ from the $\alpha$ and $\beta$-protocols in that legal users A and B have got a short key $S_A = S_B$ of length $\ell_0$ before starting the KDP. This key can be used for authentication of messages transmitted over public discussion channels in order to get finally the key of length $\ell>>\ell_0$.

In this section we consider some modification of the $\alpha'$ and $\beta'$-protocols in which instead of hash functions, extractors are used in order to generate the final key. We call these protocols the $\alpha'_{ext}$ and $\beta'_{ext}$-{\em protocols}, respectively.

\paragraph{$\alpha'_{ext}$-protocol}
Let us suppose that the users A and B have binary strings $X^k$, $Y^k$ respectively.
\begin{enumerate}
 \item The user A calculates the check string $C^r$ of length $r$ for the string $X^k$ using an error correcting $(k+r, k)$-code ${\cal C}$ that should be agreed between the legal users in advance.
 \item The user A generates a random binary string $\gamma^u$ of length $u$.
 \item The user A computes the authenticator {\bf w} for the message $(C^r,\gamma^u)$ using a keyed hash function from the $\epsilon$-ASU$_2$ class, and the key $S_A$.
 \item The user A sends to user B over a PDC the message $(C^r,\gamma^u)$ appending to it the authenticator {\bf w}.
 \item The user B verifies the authenticity of $(C^r,\gamma^u)$ using the algorithm presented in section~\ref{sc.02}.  If the authenticity of $(C^r,\gamma^u)$ is confirmed, then B goes to the next step, otherwise he rejects it.
 \item The user B corrects errors in the string $Y^k$ using the check string $C^r$. (We denote by $\tilde{Y}^k$ the string $Y^k$ after error correction).
 \item Both users A and B compute their keys as $K_A = E_{ext}(X^k,\gamma^u)$, $K_B = E_{ext}(\tilde{Y}^k,\gamma^u)$.
\end{enumerate}

\paragraph{$\beta'_{ext}$-protocol}
In a similar manner there is a modified $\beta$-protocol where the random string $\gamma^u$ is not transmitted over the PDC but it is formed from the random sequences $X^k$, $Y^k$. 

The users A and B divide each of the strings $X^k$, $Y^k$ obtained after execution of the initialization phase into two disjoint substrings $X_1^{k_1}$, $X_2^{k_2}$; $Y_1^{k_1}$, $Y_2^{k_2}$, respectively. Then they perform the following steps:
\begin{enumerate}
 \item The user A calculates the check string $C_1^{r_1}$ of length $r_1$ for the substring $X_1^{k_1}$ using an error correcting $(k_1+r_1, k_1)$-code ${\cal C}_1$.
 \item The user A calculates the check string $C_2^{r_2}$ of length $r_2$ for the substring $X_2^{k_2}$ using an error correcting $(k_2+r_2, k_2)$-code ${\cal C}_2$.
 \item The user A forms the authenticator {\bf w} for the message $(C_1^{r_1},C_2^{r_2})$, using a keyed hash function from the class $\epsilon$-ASU$_2$ and the key $S_A$ with length $\ell_0$.
 \item The user A sends to B the message $(C_1^{r_1},C_2^{r_2})$ appended with the authenticator {\bf w}.
 \item The user B verifies the authenticity of the message $(C_1^{r_1},C_2^{r_2})$ using the authentication algorithm (see section~\ref{sc.02}) and the key $S_B$. If authenticity is confirmed, then user B goes to the next step, otherwise he rejects the KDP.
 \item The user B corrects errors in the strings $Y_1^{k_1}$, $Y_2^{k_2}$ using the check strings $C_1^{r_1}$ and $C_2^{r_2}$. (We denote by $\tilde{Y}_1^{k_1}$, $\tilde{Y}_2^{k_2}$ the strings $Y_1^{k_1}$, $Y_2^{k_2}$ after error correction.)
 \item The user A takes the string $X_2^{k_2}$ as seed $\gamma^u$ and the user B takes the string $\tilde{Y}_2^{k_2}$ as seed $\gamma^u$.
 \item Both users A and B compute their keys as $K_A = E_{ext}(X_1^{k_1},X_2^{k_2})$, $K_B = E_{ext}(\tilde{Y}_1^{k_1},\tilde{Y}_2^{k_2})$.
\end{enumerate}
Let us estimate the key rate of these protocols.

\begin{thm}\label{th.15} Let us suppose that the users A, B and the adversary E have binary strings $X^k$, $Y^k$ and $Z^k$, respectively after execution of the initialization phase over the wire-tape channel, $p_m=\pr{x_i\not=y_i}$, $p_w=\min(\pr{x_i\not=z_i},\pr{y_i\not=z_i}$, $p_m\geq 0$, $p_w>p_m$. We assume that the users A and B share initially a short key $S$ of length $\ell_0$ in order to authenticate messages transmitted over the PDC. 

Then A and B are able to form a common key of length $\ell$ satisfying the requirements~(\ref{eq.003})-(\ref{eq.007}) after the execution of the $\alpha'_{ext}$ and $\beta'_{ext}$-protocols if the lengths $k_1$, $k_2$ of substrings $X^k$, $Y^k$ and $Z^k$ and $\ell_0$ satisfy the equations listed below: 
\begin{eqnarray}
k_1 &=& -\frac{\log P^{adm}_e}{E(R_{c_1})}, \label{eq.094} \\
k_1\cdot H_{\infty} &=& \ell c + r_1 - \log P^{adm}_{risk} + u \nonumber \\
 & & + 3 \log\ell\ \left(\frac{I^{adm}}{2\ell}\right)^{-2} + 3
, \label{eq.095}
\end{eqnarray}
where $R_{c_1}$ and $E(R_{c_1})$ are calculated by~(\ref{eq.013})-(\ref{eq.015}) and
\begin{equation}
k_2 = \left\{\begin{array}{ll}
0 & \mbox{for the $\alpha'_{ext}$-protocol,} \\
u & \mbox{for the $\beta'_{ext}$-protocol,} 
\end{array}\right. \label{eq.096}
\end{equation}
and
\begin{equation}
u = \left\lceil\frac{\left\lceil\log k_1\,\left(\frac{I^{adm}}{2\ell}\right)^{-2}\right\rceil}{\ln c}\right\rceil\cdot\left\lceil\log k_1\,\left(\frac{I^{adm}}{2\ell}\right)^{-2}\right\rceil \label{eq.097}
\end{equation}
is the number of the extractor random symbols, $c$ is the parameter under optimization,
\begin{eqnarray}
\ell_0 &=& \frac{a(2+i)}{2^i}, \label{eq.098} \\
\frac{i+1}{2^{\frac{a}{2^i}}} &\leq& P_d^{adm}, \label{eq.099} 
\end{eqnarray}
where
\begin{equation}
a = \left\{\begin{array}{ll}
r_1+u & \mbox{for the $\alpha'_{ext}$-protocol,} \\
r_1+r_2 & \mbox{for the $\beta'_{ext}$-protocol,} 
\end{array}\right. \label{eq.100}
\end{equation}
and $r_2$ being the number of check symbols of the error correcting $(k_2+r_2,k_2)$-code ${\cal C}_2$ found similarly as in eq's~(\ref{eq.094})-(\ref{eq.095}). The key rate is then determined as: 
\begin{equation}
R_{\alpha'_{ext}} = \max_c\frac{\ell}{k_1} \ \ ,\ \  R_{\beta'_{ext}} = \max_c\frac{\ell}{k_1+k_2} . \label{eq.101}
\end{equation}
\end{thm}

{\em Proof.}
The relations~(\ref{eq.094}),~(\ref{eq.095}) and~(\ref{eq.097}) can be proved similarly as~(\ref{eq.063}),~(\ref{eq.064}) and~(\ref{eq.065}) in theorem~\ref{th.11}. The relation~(\ref{eq.096}) is apparent from the protocols description. In order to prove~(\ref{eq.098}),~(\ref{eq.099}) we assume that for authentication of messages of length $a$ (see relation~(\ref{eq.100})) an $\epsilon$-ASU$_2$-hash-function is used. Relying on~(\ref{eq.017}) we write $2^a = q^{2^i}$, $2^{\ell_0}=q^{i+2}$, $\epsilon=\frac{i+1}{q}$. Let us put $q=2^b$, then 
\begin{equation}
a=2^ib\ ,\ \ell_0 = b(i+2)\ ,\ \epsilon=\frac{i+1}{2^b}. \label{eq.102}
\end{equation}
Let us assume that the probability of false message deception is equal to $\epsilon=P_d^{adm}$. Then from~(\ref{eq.102}), the relations~(\ref{eq.098}),~(\ref{eq.099}) are valid. The relation~(\ref{eq.101}) follows from the protocols definition taking into account that the number of the extractor random bits can be optimized with respect to $c$.  \hfill $\Box$

By substituting~(\ref{eq.095}) into~(\ref{eq.101}) and using~(\ref{eq.016}) we get that as $\ell\to\infty$:
\begin{eqnarray}
R_{\alpha'_{ext}} &=& \frac{\ell\left[H_{\infty}-g(p_m)\right]}
{Den_1}, \label{eq.103} \\
R_{\beta'_{ext}} &=& \frac{\ell\left[H_{\infty}-g(p_m)\right]}
{Den_2}, \label{eq.104}
\end{eqnarray}
where
\begin{eqnarray*}
Den_1 &=& \ell c - \log P_{risk} + u + 3 \log\ell\ \left(\frac{I^{adm}}{2\ell}\right)^{-2} + 3, \\
Den_2 &=& Den_1 + u. 
\end{eqnarray*}
From~(\ref{eq.103}),~(\ref{eq.104}), we have $R_{\alpha'_{ext}}\geq R_{\beta'_{ext}}$. When $\ell\to\infty$ both protocols have the same key rates 
\begin{equation}
R_{\alpha'_{ext}} = R_{\beta'_{ext}} = H_{\infty}-g(p_m). \label{eq.105}
\end{equation}
Let us compare the key rates of the $\alpha'_{ext}$ and $\beta'_{ext}$-protocols and the $\alpha'$ and $\beta'$-protocols. In~\cite{xxvi} the following relations have been proved: 
$$\left.\begin{array}{rcl}
R_{\alpha'} &\to& H_2(p_w)-g(p_m) \\
R_{\beta'} &\to& \frac{1}{2}\left[H_2(p_w)-g(p_m)\right]
\end{array}\right\} \mbox{ as }\ell\to\infty.$$
Comparing these relations with~(\ref{eq.105}) we may conclude that $R_{\alpha'} \geq R_{\alpha'_{ext}} = R_{\beta'_{ext}}$ for any values $p_w$ and $p_m$. $R_{\beta'}$ can be either larger or smaller than $R_{\alpha'_{ext}} = R_{\beta'_{ext}}$ depending on the ratio of $p_w$ and $p_m$.

\begin{figure}[!t]
\centering
 \includegraphics[width=3in]{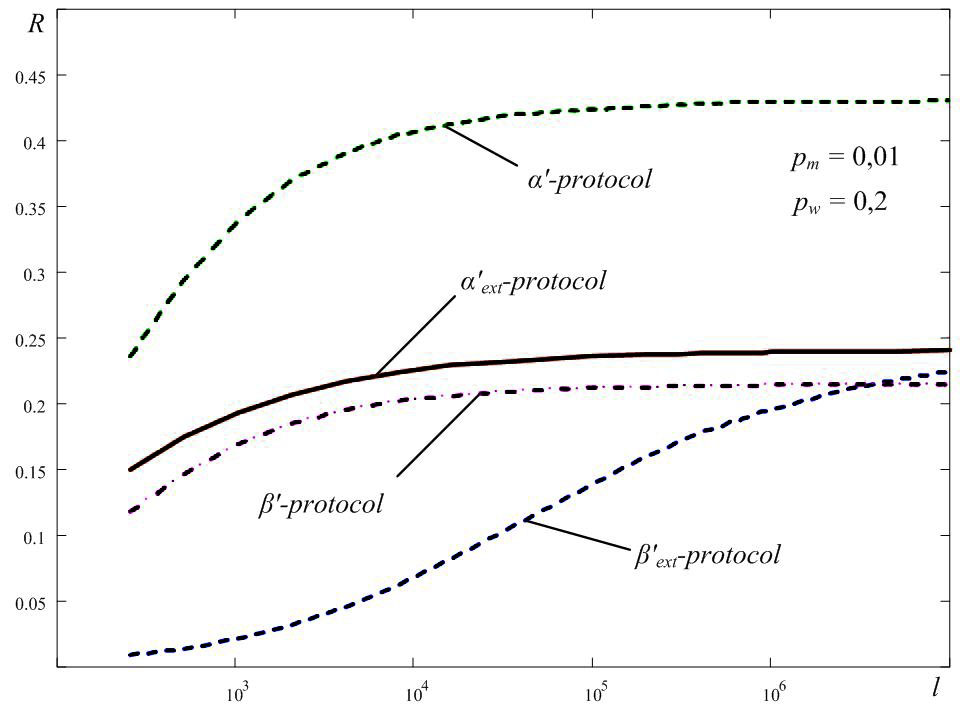} 
\caption{Key rates versus its length for different KDP provided initially with short authentication keys.}
\label{fig.06}
\end{figure}

In order to illustrate the above assertions we plot in figure~\ref{fig.06} the dependence of the key rate versus its length for the $\alpha'$, $\beta'$, $\alpha'_{ext}$ and $\beta'_{ext}$-protocols, given $p_w = 0.2$, $p_m = 0.01$ and $I^{adm} = 10^{-30}$, $P_e^{adm} = P_d^{adm} = P_{risk}^{adm} = 10^{-5}$.

\section{Two-stage (hybrid) protocols with extractors} \label{sc.05}

We remember that the hybrid protocols~\cite{xxvi,xxxviii} are combinations of protocol pairs $(\alpha,\alpha')$, $(\alpha,\beta')$, $(\beta,\alpha')$, $(\beta,\beta')$ where the first protocol in each pair is used in order to generate a relatively short key $S$ of length $\ell_0$ required for hash function and check bits authentication, whereas the second protocol is used in order to form the final key $K$.

The keys $S$ and $K$ can be obtained by execution of the privacy amplification procedure based either on the use of hash functions or extractors. This means that for every above mentioned hybrid protocol pair, there are four variants of hashing or extracting applications. In total, there can be formed 16 protocols, which in turn can be split into four groups as shown in table~\ref{tb.01}.
\begin{table}[!t]
\caption{Groups of hybrid protocols}
\centering
\begin{tabular}{||c|p{3.5cm}|p{1.5cm}|p{1.5cm}||} \hline \hline
 \multicolumn{2}{||c|}{Protocol group} & Authentication key generation & Key generation \\ \hline
1 & $(\alpha,\alpha')$, $(\alpha,\beta')$, $(\beta,\alpha')$, $(\beta,\beta')$ & hashing & hashing \\ \hline
2 & $(\alpha,\alpha'_{ext})$, $(\alpha,\beta'_{ext})$, $(\beta,\alpha'_{ext})$, $(\beta,\beta'_{ext})$ & hashing & extracting \\ \hline
3 & $(\alpha_{ext},\alpha')$, $(\alpha_{ext},\beta')$, $(\beta_{ext},\alpha')$, $(\beta_{ext},\beta')$ & extracting & hashing \\ \hline
4 & $(\alpha_{ext},\alpha'_{ext})$, $(\alpha_{ext},\beta'_{ext})$, $(\beta_{ext},\alpha'_{ext})$, $(\beta_{ext},\beta'_{ext})$ & extracting & extracting \\ \hline \hline
\end{tabular}
\label{tb.01}
\end{table} 

The first group of protocols was investigated in~\cite{xxvi}, and there it has been proved that every such protocol can be the most efficient depending on the additional key requirements imposed to it. It is worth to note that even for large length $\ell$ of the key $K$, the length $\ell_0$ of the authentication key occurs moderate~\cite{xxvi} (p. 2543). If $\ell=32000$ ($p_m = 0.01$, $p_w = 0.2$, $P^{adm} = 5\cdot 10^{-6}$), then $\ell_0 = 678$. But since, as shown in section~\ref{sc.03}, extractors are superior than hash functions only with large key lengths, their application is useless in the first stage of the hybrid protocols, where a short key is required.

Therefore, the protocols from groups 3 and 4 have not been considered. It is sufficient to investigate protocols from the second group, where the authentication key is generated by hashing and the generation of the final keys is performed by extraction. Thus we consider the following hybrid protocols: $(\alpha,\alpha'_{ext})$, $(\alpha,\beta'_{ext})$, $(\beta,\alpha'_{ext})$, $(\beta,\beta'_{ext})$. For a more detailed description with the design of these protocols and the specification requirements of each protocol component, we refer to~\cite{xxvi} (p. 2544).

\begin{figure}[!t]
\centering
 \includegraphics[width=3.3in]{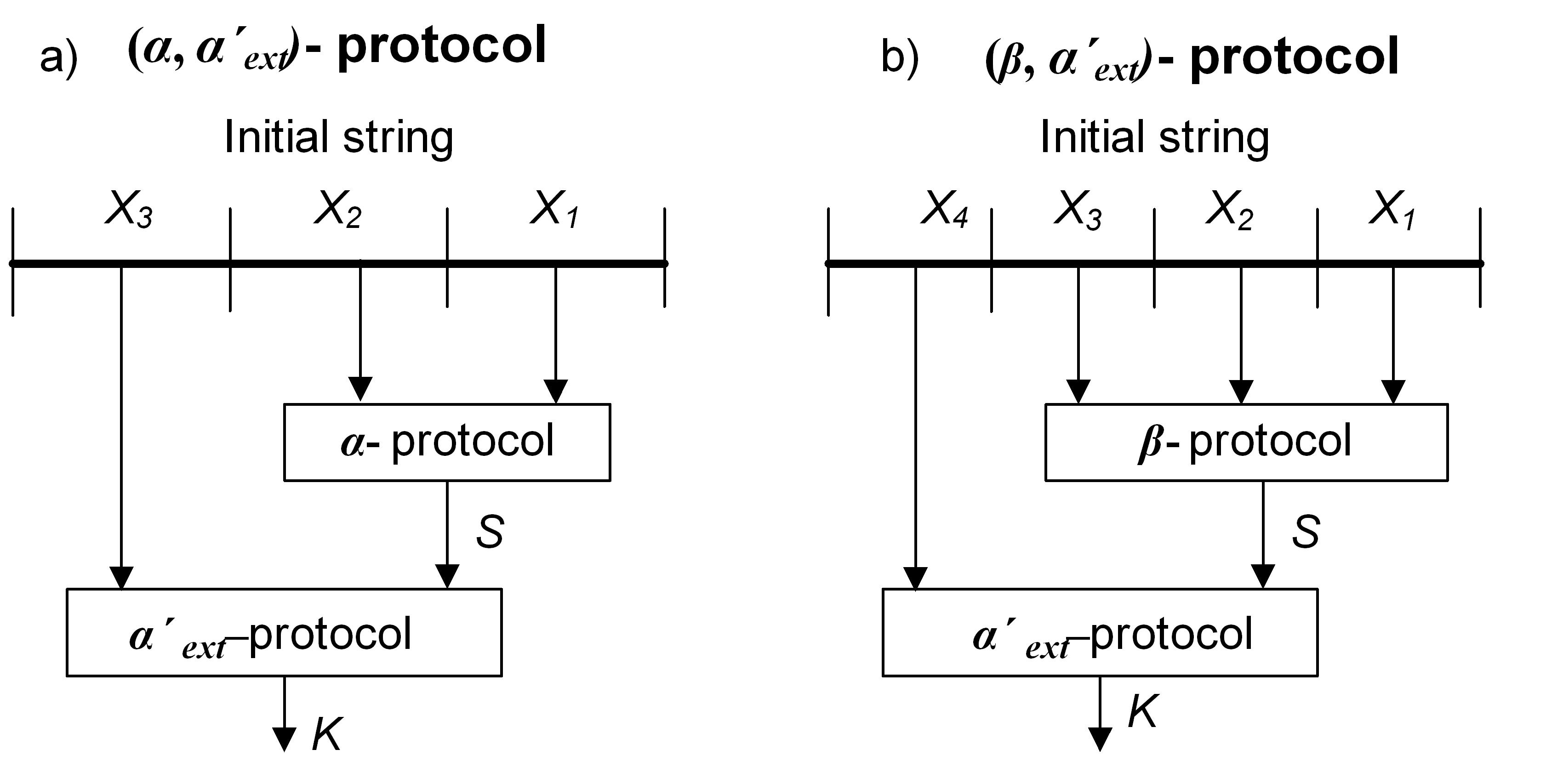} \\
 \includegraphics[width=3.3in]{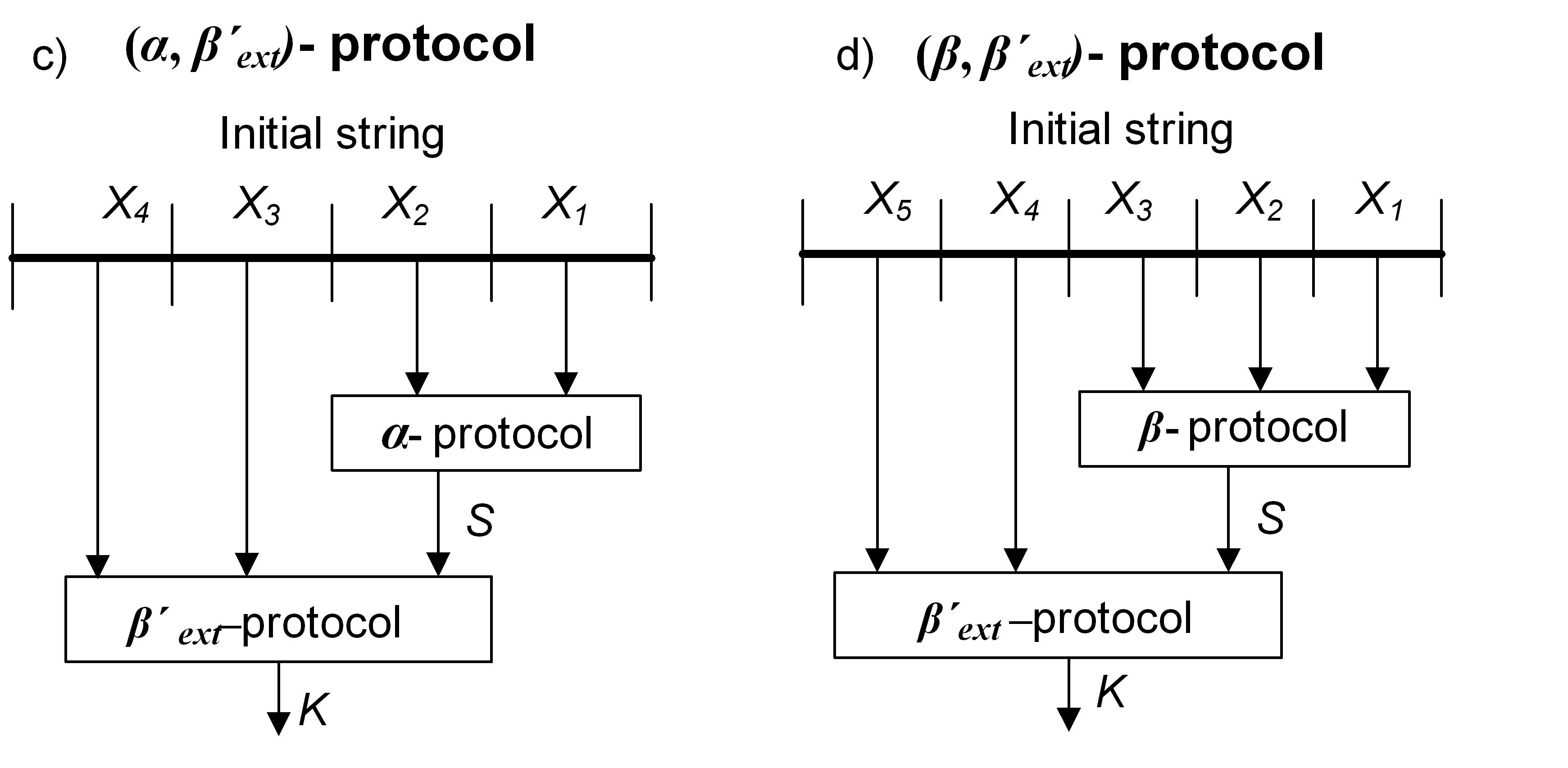} \\
\caption{Different types of hybrid protocol with extractors.}
\label{fig.07}
\end{figure}

Let us give a short description of the $(\alpha,\alpha'_{ext})$-protocol. It is based on the $(\alpha,\alpha')$-protocol proposed by Korzhik and Morales~\cite{xxxviii}. In this protocol, the sequences $X^k$, $Y^k$ of users A and B are divided into three parts $X_1^{k_1}$, $X_2^{k_2}$, $X_3^{k_3}$ and $Y_1^{k_1}$, $Y_2^{k_2}$, $Y_3^{k_3}$ respectively (see figure~\ref{fig.07}-a). The subsequences $X_1^{k_1}$, $X_2^{k_2}$, ($Y_1^{k_1}$, $Y_2^{k_2}$) are used for the generation of the authentication keys $S_A$ ($S_B$). The subsequence $X_3^{k_3}$ ($Y_3^{k_3}$) and the key $S_A$ ($S_B$) are used in the $\alpha'_{ext}$-protocol for final $K_A$ ($K_B$) key generation. The key rate of this protocol can be written as
\begin{equation}
R_{(\alpha,\alpha'_{ext})} = \frac{\ell}{k_1+k_2+k_3} = \frac{\ell}{k_3+\frac{\ell_0}{R_{\alpha}}}, \label{eq.106}
\end{equation}
where $R_{\alpha} = \frac{\ell}{k_1+k_2}$ is the key $S$ rate at the length $\ell_0$ in the $\alpha$-protocol.

Let us prove the following lemma.

\begin{lemma}\label{lm.03}
 The convergence $\frac{\ell_0}{\ell}\to 0$, as $\ell\to\infty$, holds in the $(\alpha,\alpha'_{ext})$-protocol.
\end{lemma}

{\em Proof.} Let us consider the $(\alpha,\alpha'_{ext})$-protocol. According to~(\ref{eq.100}) the input block length of the $\epsilon$-$ASU_2$ hash function used in the $\alpha_{ext}$-protocol is equal to $a=r_3+u$. We show initially that $r_3+u\to\infty$ as $\ell\to\infty$. The relations~(\ref{eq.097}) and~(\ref{eq.014}) give
\begin{eqnarray}
u &\approx& \frac{\left(\log k_3\,\left(\frac{I^{adm}}{2\ell}\right)^{-2}\right)^2}{\ln c}, \label{eq.106a} \\
r_3 &=& (1-R_c)k_3. \label{eq.106b} 
\end{eqnarray}
Thus in order to prove that $r_3+u\to\infty$ it is necessary to show that $k_3\to\infty$ as $\ell\to\infty$. Taking into account equation~(\ref{eq.016}), we can present~(\ref{eq.095}) in the form
\begin{equation}
k_3 = \frac
{\ell c -\log P^{adm}_{risk} + u +3 \log\ell \left(\frac{I^{adm}}{2\ell}\right)^{-2}   + 3}
{H_{\infty}(p_w) - g(p_m)}. \label{eq.107}
\end{equation}
From the above relation, it follows that $k_3\to\infty$ as $\ell\to\infty$ and both~(\ref{eq.106a}),~(\ref{eq.106b}) result as $r_3\to\infty$ and $u\to\infty$. Then using~(\ref{eq.098}) for an estimation of the hash function parameters, we may write 
$$
\lim_{i\to\infty}\frac{\ell_0}{a} = \lim_{i\to\infty}\frac{i+2}{2^i} = 0. 
$$
\hfill $\Box$

Using~(\ref{eq.107}), the relation~(\ref{eq.106}) can be expressed as
\begin{equation}
R_{(\alpha,\alpha'_{ext})} = \frac{H_{\infty}(p_w) - g(p_m)}{Den_3}, \label{eq.109}
\end{equation}
where
\begin{eqnarray*}
Den_3 &=& c  \\
 & & -\ell^{-1}\left[\log P^{adm}_{risk} + u +3 \log\ell \left(\frac{I^{adm}}{2\ell}\right)^{-2}  + 3\right]  \\
 & & + \frac{\ell_0}{\ell}\frac{H_{\infty}(p_w) - g(p_m)}{R_{\alpha}}. 
\end{eqnarray*}
Taking into account that $R_{\alpha}$ is constant and $\frac{\ell_0}{\ell}\to 0$ (see Lemma~\ref{lm.03})
we can see that the last term in the denominator $Den_3$ of~(\ref{eq.109}) approaches to zero as $\ell\to\infty$. The other terms in the denominator $Den_3$ also approach to zero because they consist either of values approaching zero or have a logarithmic dependence on $\ell$.

Since the right side of~(\ref{eq.109}) approaches a maximum, $R_{(\alpha,\alpha'_{ext})}$ approaches a maximum as $c\to 1$, then the following asymptotic estimation holds for the key rate of the $(\alpha,\alpha'_{ext})$-protocol
\begin{eqnarray}
 & & R_{(\alpha,\alpha'_{ext})} = H_{\infty}(p_w) - g(p_m), \label{eq.110a} \\
 & & R_{(\alpha,\alpha'_{ext})}^* = H_{\infty}(p_w) \mbox{ as }p_m\to 0. \label{eq.110}
\end{eqnarray}
In the $(\beta,\alpha'_{ext})$-protocol (see figure~\ref{fig.07}-b) the sequences $X^k$, $Y^k$ are divided into four parts $X_1^{k_1}$, $X_2^{k_2}$, $X_3^{k_3}$, $X_4^{k_4}$ and $Y_1^{k_1}$, $Y_2^{k_2}$, $Y_3^{k_3}$, $Y_4^{k_4}$, respectively. The subsequences $X_1^{k_1}$, $X_2^{k_2}$, $X_3^{k_3}$ and $Y_1^{k_1}$, $Y_2^{k_2}$, $Y_3^{k_3}$ are used in the $\beta$-protocol in order to generate the authentication key $S$ of length $\ell_0$. The subsequence $X_4^{k_4}$ ($Y_4^{k_4}$) is used in the $\alpha'_{ext}$-protocol for final keys $K_A$, $K_B$ generation. One can write
$$
R_{(\beta,\alpha'_{ext})} = \frac{\ell}{k_4+\frac{\ell_0}{R_{\beta}}}, 
$$
where $R_{\beta} = \frac{\ell_0}{k_1+k_2+k_3}$ is the authentication key rate.

By comparing this protocol with the previous one, we can conclude that for the same length $\ell$ of the final key, the equality $k_4 = k_3$ should hold. As it was shown in~\cite{xxvi}, $R_{\beta}\geq R_{\alpha}$ and the length $\ell_0$ of authentication key for the $\alpha$-protocol is larger than the length $\ell_0$ of the authentication key for the $\beta$-protocol. Hence
$$\frac{\ell_0(\alpha\mbox{-protocol})}{R_{\alpha}} \geq \frac{\ell_0(\beta\mbox{-protocol})}{R_{\beta}}$$
and $R_{(\beta,\alpha'_{ext})} \geq R_{(\alpha,\alpha'_{ext})}$.

It is easy to show that
\begin{equation}
R_{(\beta,\alpha'_{ext})} = H_{\infty}(p_w)- g(p_m) \mbox{ as }\ell\to \infty ,\label{eq.112}
\end{equation}
that coincides with the key rate of the $(\alpha,\alpha'_{ext})$-protocol, see~(\ref{eq.110}). If $p_m = 0$ we get by~(\ref{eq.112})
$$
R_{(\beta,\alpha'_{ext})}^* = H_{\infty}(p_m) . 
$$

Next let us consider the $(\alpha,\beta'_{ext})$-protocol (see figure~\ref{fig.07}-c) in which each sequence $X^k$, $Y^k$ is divided into four parts $X_1^{k_1}$, $X_2^{k_2}$, $X_3^{k_3}$, $X_4^{k_4}$ and $Y_1^{k_1}$, $Y_2^{k_2}$, $Y_3^{k_3}$, $Y_4^{k_4}$, respectively.  The subsequences $X_1^{k_1}$, $X_2^{k_2}$ and $Y_1^{k_1}$, $Y_2^{k_2}$ are used in the $\alpha$-protocol for the generation of the key $S_A$, or $S_B$ for the $\beta'_{ext}$-protocol, while subsequences $X_3^{k_3}$, $X_4^{k_4}$ and $Y_3^{k_3}$, $Y_4^{k_4}$ are used in the $\beta'_{ext}$-protocol for the final key $K_A$ ($K_B$) generation assuming $X_4^{k_4}$ and $Y_4^{k_4}$ as random ``seeds'' $\gamma^{k_4}$ while using in extractor.

Similarly to~(\ref{eq.106}) we can write
$$
R_{(\alpha,\beta'_{ext})} = \frac{\ell}{k_3+u+\frac{\ell_0}{R_{\alpha}}}. 
$$
The parameter $\ell_0$ is smaller in this protocol than in the $(\alpha,\alpha'_{ext})$-protocol, because only to authenticate the check sequences $X_3^{k_3}$ and $X_4^{k_4}$ of total length $r_1 + r_2$ there is used an $\epsilon$-$ASU_2$ hash-function. However the additional item ($u$) calculated by~(\ref{eq.097}) increases the denominator and hence $R_{(\alpha,\beta'_{ext})}\leq R_{(\alpha,\alpha'_{ext})}$.

In the $(\beta,\beta'_{ext})$-protocol, each sequence $X^k$, $Y^k$ is divided into five parts $X_1^{k_1}$, $X_2^{k_2}$, $X_3^{k_3}$, $X_4^{k_4}$, $X_5^{k_5}$ and $Y_1^{k_1}$, $Y_2^{k_2}$, $Y_3^{k_3}$, $Y_4^{k_4}$ $Y_5^{k_5}$, respectively (see figure~\ref{fig.07}-d). The subsequences $X_1^{k_1}$, $X_2^{k_2}$, $X_3^{k_3}$ and $Y_1^{k_1}$, $Y_2^{k_2}$, $Y_3^{k_3}$ are used in the $\beta$-protocol to generate the keys $S_A$, $S_B$. The subsequences $X_4^{k_4}$, $X_5^{k_5}$ and $Y_4^{k_4}$ $Y_5^{k_5}$ are used in the $\beta'_{ext}$-protocol to generate the keys $K_A$, $K_B$. Let us write
$$
R_{(\beta,\beta'_{ext})} = \frac{\ell}{k_4+u+\frac{\ell_0}{R_{\beta}}}. 
$$
In this relation by the same reason mentioned during the analysis of the $(\alpha,\beta'_{ext})$-protocol, the value $\ell_0$ will be smaller than in the $(\beta,\alpha'_{ext})$-protocol and the rate $R_{\beta}$ is larger. Therefore the third item is slightly decreasing. However the presence of sufficiently large item u results in a key rate decreasing giving the inequality $R_{(\beta,\beta'_{ext})}\leq R_{(\beta,\alpha'_{ext})}$.

There may be for the $(\alpha,\beta'_{ext})$, $(\beta,\alpha'_{ext})$-protocols some equivalent statement to Lemma~\ref{lm.03}, e.g. it can be proved that $\frac{\ell_0}{\ell}\to 0$ as $\ell\to\infty$. Furthermore, by writing the relations for $k_3$, $k_4$ and $u$, from theorem~\ref{th.13} it is very simple to get asymptotically achievable the key rate for the $(\alpha,\beta'_{ext})$, $(\beta,\alpha'_{ext})$-protocols.
\begin{equation}
R_{(\alpha,\beta_{ext})} = R_{(\beta,\beta'_{ext})} = H_{\infty}(p_w) - g(p_m).
 \label{eq.116}
\end{equation}
Comparing~(\ref{eq.110a}),~(\ref{eq.112}) and~(\ref{eq.116}) we can see that asymptotically all hybrid protocols have the same key rates, however the $(\beta,\alpha'_{ext})$-protocol has non-asymptotically the largest key rate among all above considered hybrid protocols.

It is worth to compare this protocol with the $(\beta,\alpha')$-protocol, that (as shown in~\cite{xxvi}) has maximum possible key rate for sufficiently large $\ell$ among all hybrid protocols using hash functions in the privacy amplification procedure.

For the $(\beta,\alpha')$-protocol one can write~\cite{xxvi}
\begin{equation}
R_{(\beta,\alpha')} = H_2(p_m) - g(p_m)
 \label{eq.117}
\end{equation}
By comparing~(\ref{eq.112}) with~(\ref{eq.117}) and taking into account that $H_2(p_m)\geq H_{\infty}(p_m)$ we can see that for large $\ell$, $R_{(\beta,\alpha'_{ext})}\geq R_{(\beta,\alpha'_{ext})}$.

From this inequality, it follows that an implementation of extractors for large key length in hybrid protocols are inefficient.

\begin{figure}[!t]
\centering
 \includegraphics[width=3.3in]{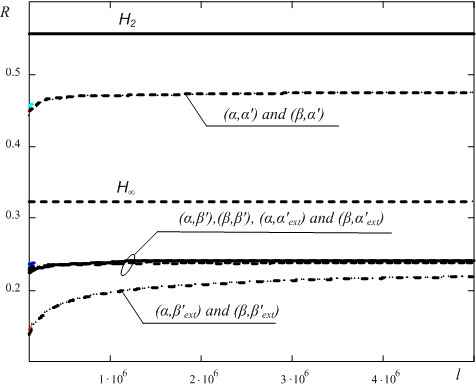} 
\caption{The key rates for hybrid protocol.}
\label{fig.08}
\end{figure}

In figure~\ref{fig.08} there are plotted the key rates versus its length for hybrid protocols under the conditions $p_m = 0.01$, $p_w = 0.2$, $I^{adm} = 10^{-30}$,  $P_e^{adm} = P_d^{adm} = P_f^{adm} = 10^{-5}$.

The curves $R(\ell)$ were plotted with the use of the technique proposed in~\cite{xxvi}. They clearly demonstrate a behavior of the key rate depending on the key length for different protocols. We can see that the $(\alpha,\alpha')$ and $(\beta,\alpha')$-protocols have the greatest key rates among all hybrid protocols. The $(\alpha,\beta')$, $(\beta,\beta')$, $(\alpha,\alpha'_{ext})$, and $(\beta,\alpha'_{ext})$-protocols have approximately equal key rates for large key length and the $(\alpha,\beta'_{ext})$ and $(\beta,\beta'_{ext})$-protocols have the least key rate among the above considered hybrid protocols.

\section{Conclusions} \label{sc.06}

In the current paper, an investigation of key distribution protocols based on noisy channels started in~\cite{xxvi} has been continued with such a difference that extractors are used instead of hash-functions in the privacy amplification procedure. The main goal was to prove extractor-based protocols efficiency by the criterion of key rate maximization. The relations are non-asymptotic and constructive because they do not include some uncertain coefficients in notations, in contrast with other papers.

We use the modified Trevisan's extractor~\cite{xxxvii,xxxvi} in our paper. It has been proposed new $\beta_{ext}$ and $\beta_{ext}^*$-protocols which differ from those known before~\cite{xxi} because the extractor's seed is not transmitted over the PDC but, instead, it is generated from random sequences obtained by legal user after the execution of the initialization phase. We proved that the use of extractors in the $\alpha_{ext}$ and $\beta_{ext}$-protocols increases the rate, in comparison with hashing-based protocols only for very large key length $\ell$ (typically $\ell\in[10^5, 10^6]$) and for some specified values of the error probabilities both in the main and in the wire-tap channels.

It was investigated a performance evaluation of the so called extractor-based hybrid protocols, consisting of two protocols executed in a serial manner where the first protocol in a pair is used for the generation of a relatively short key $S$ of length $\ell_0$. This key is necessary for authentication of check bits, and a random number (seed) of extractor. The second protocol is used for the final key generation. We prove that extractor based protocols should be used only in the second protocol of the pair.

We selected four hybrid protocols for further investigation $(\alpha,\alpha'_{ext})$, $(\beta,\alpha'_{ext})$, $(\alpha,\beta'_{ext})$ and $(\beta,\beta'_{ext})$-protocols. The relations for their key rates have been derived for both finite and asymptotically growing key lengths. The greatest key rate is got for the $(\beta,\alpha'_{ext})$-protocol. This protocol was compared with the $(\beta,\alpha')$-protocol considered in~\cite{xxvi}, which has the greatest key rate among all hybrid hashing-based protocols. The investigations showed that hybrid protocols with the use of extractor-based second stage protocols are less efficient than hashing-based protocol.

We investigated also (but not in a deeper detail) other variants of extractors from~\cite{xxxvi}. Even with some improvement of their characteristics (in the sense of the seed length), the general conclusion is kept the same: the use of extractors is justified only with very large key length.

We get also asymptotic estimates for the key rates of all proposed protocols that allows to compare the potential efficiency of all considered early protocols. These relations are presented in table~\ref{tb.02}. We can see that asymptotically all hybrid protocols have the same key rates equal to $H_{\infty}(p_w) - g(p_m)$, that is larger than the key rates for single $\alpha_{ext}$ and $\beta_{ext}$-protocols that is equal to $\frac{H_{\infty}(p_w) - g(p_m)}{1 + 2 g(p_m)}$.

These relations are similar ``on structure'' to relations for key capacity $g(p_w) - g(p_m)$,~\cite{xxiii}, but differ from the last one in changing of Shannon's entropy $g(p_w)$ to min entropy $H_{\infty}(p_w)$.

If the main channel is noiseless then all protocols using extractors have the same asymptotic key rates equal to $H_{\infty}(p_w)$. It is worth to note that asymptotically all hybrid extractor-based protocols are inferior to hash-based protocols. But this conclusion may be considered as a consequence of crude estimate of information leaking to eavesdropper based on the use of min entropy.

We summarize the key rates for different KDP in table~\ref{tb.02}. It can be seen from this table how closer or farther are the key rates to the secret key capacity given by~(\ref{eq.009}).

\begin{table}
 \caption{Key rates for different KDP} \label{tb.02}
\centering
\begin{tabular}{|c|c|c|c|} \hline 
 Protocol type & (2) & (3) & (4) \\ \hline
 $\alpha_{ext}$ & $\frac{\ell}{\frac{T_2}{T_1}(1+2g(p_m)) + 2u + 2r_0}$ & $\frac{T_1}{1+2g(p_m)}$ & $T_0$ \\ \hline
 $\beta_{ext}$ & $\frac{\ell}{\left(\frac{T_2}{T_1}+u\right)(1+2g(p_m)) + 2r_0}$ & $\frac{T_1}{1+2g(p_m)}$ & $T_0$ \\ \hline
 $\alpha'_{ext}$ & $\ell \frac{T_1}{T_2}$ & $T_1$ & $T_0$ \\ \hline
 $\beta'_{ext}$ & $\ell \frac{T_1}{T_2}$ & $T_1$ & $T_0$ \\ \hline
 $(\alpha,\alpha'_{ext})$ & $\ell \frac{T_1}{T_2+\ell_0 \frac{T_1}{R_{\alpha}}}$ & $T_1$ & $T_0$ \\ \hline
 $(\beta,\alpha'_{ext})$ & $\ell \frac{T_1}{T_2+\ell_0 \frac{T_1}{R_{\beta}}}$ & $T_1$ & $T_0$ \\ \hline
 $(\alpha,\beta'_{ext})$ & $\ell \frac{T_1}{T_2+u+\ell_0 \frac{T_1}{R_{\alpha}}}$ & $T_1$ & $T_0$ \\ \hline
 $(\beta,\beta'_{ext})$ & $\ell \frac{T_1}{T_2+u+\ell_0 \frac{T_1}{R_{\beta}}}$ & $T_1$ & $T_0$ \\ \hline
\end{tabular} \vspace{2ex}\\ 
\begin{minipage}{7cm}
 (2): Non asymptotic relations for key rates $R_k$ \\
 (3): Asymptotic relations for $R_k$, as $\ell \to\infty$ \\
 (4): Asymptotic relations for $R_k$, as $\ell \to\infty$, $p_m\to 0$ \\
\end{minipage}
\begin{eqnarray*}
T_0 &=& H_{\infty}(p_w) \\
T_1 &=& T_0 - g(p_m) \\
T_2 &=& c\ell + u + a \\
a  &=& -2 P^{adm}_{risk} + 3 \log\ell\ \left(\frac{I^{adm}}{2\ell}\right)^{-2} + 3
\end{eqnarray*}
$u$ is given by relation~(\ref{eq.066}) and $R_{\alpha}$, $R_{\beta}$ were introduced in~\cite{xxvi}.
\end{table}


\bibliographystyle{IEEEtran}
\bibliography{10ykm}

\end{document}